\PassOptionsToPackage{hyperfootnotes=false}{hyperref}
\documentclass[fleqn,usenatbib]{mnras}

\usepackage{newtxtext,newtxmath}

\usepackage{todonotes}
\usepackage{adjustbox}
\usepackage{csquotes}
\usepackage[T1]{fontenc}
\hypersetup{hypertexnames=false}

\DeclareRobustCommand{\VAN}[3]{#2}
\let\VANthebibliography\thebibliography
\def\thebibliography{\DeclareRobustCommand{\VAN}[3]{##3}\VANthebibliography}

\usepackage{graphicx}	% Including figure files
\usepackage{amsmath}	% Advanced maths commands

\title[Analytic DF for DM Haloes with Symbolic Regression]{A Simple Analytic Distribution Function for Cosmological Dark Matter Haloes from Symbolic Regression}

\author[Maxim Botbol et al.]{
Maxim Botbol,$^{1,2}$\thanks{E-mail: mrb209@cam.ac.uk}
Vasily Belokurov,$^{2}$
N. Wyn Evans,$^{2}$
Pei Ting Zeng$^{3}$
\\
$^{1}$École Polytechnique, Institut Polytechnique de Paris, Palaiseau, France\\
$^{2}$Institute of Astronomy, Madingley Rd, Cambridge, CB3 0HA\\
$^{3}$Department of Computational and Applied Mathematics, The University of Chicago, Chicago, IL 60637 USA\\}
\date{Accepted XXX. Received YYY; in original form ZZZ}
\pubyear{\the\year{}}

\begin{document}
\label{firstpage}
\pagerange{\pageref{firstpage}--\pageref{lastpage}}
\maketitle

\begin{abstract}
We develop a pipeline to extract and symbolically regress empirical distribution functions for 30 Milky Way-mass dark matter only (DMO) haloes, from N-body cosmological simulations under the spherical, isotropic assumption. Using similarity scaling and symbolic regression, implemented by PySR, we identify physically admissible analytic expressions that are common to the Pareto fronts of all cross-validation folds and discuss their physical consequences. The surviving expressions, passing a loss cut, belong to the simple family
$f\propto\hat\varepsilon^{\nu}/(1-\hat\varepsilon)^{\mu}$ and its simple extensions; the simplest of them implies an extrapolated inner density slope
of $0.92$ and a slope of $3.2$ at the virial radius, compared with $1$ and $2.9$ for NFW, and all of them closely agree with NFW and Einasto profiles within the virial radius.
\end{abstract}


\section{Introduction}
In the $\Lambda$CDM paradigm, dark-matter haloes assemble hierarchically through mergers and accretion and provide the gravitational scaffolding on which galaxies form. Navarro, Frenk, and White showed that, despite the diversity of these assembly histories, the spherically averaged density profiles of relaxed haloes could be described approximately by a common radial form with a scale radius, $r_s$, and a density normalisation, $\rho_s$ \citep{Navarro1997NFW}. The NFW profile steepens from an inner $r^{-1}$ cusp to an outer $r^{-3}$ decline, passing through logarithmic slope $-2$ at $r_s$. Because the outer decline gives a logarithmically divergent total mass, a finite halo mass must be defined within a conventional boundary. The virial radius, $R_{\rm vir}$, encloses a specified mean overdensity relative to a reference cosmological density, and the corresponding virial mass is $M_{\rm vir}=M(<R_{\rm vir})$ \citep{White2001HaloMass}. The concentration $c_{\rm vir}\equiv R_{\rm vir}/r_s$ measures the separation between the halo boundary and the scale radius. Once the overdensity convention and redshift are fixed, $(r_s,\rho_s)$ and $(M_{\rm vir},c_{\rm vir})$ are equivalent two-parameter descriptions of an NFW halo, and both parameterisations are commonly used.

Halo mass and concentration are not statistically independent. Simulations find a mass--concentration relation that evolves with redshift and depends on cosmology, while much of its scatter at fixed mass is connected to differences in halo assembly history \citep{Bullock2001Profiles,Wechsler2002Concentrations,Ludlow2014MassConcentration}. Later calculations extended the approximate regularity of halo density profiles over nearly 20 orders of magnitude in halo mass \citep{Wang2020Halostructure}. The same compact radial form thus describes objects whose masses and assembly histories differ enormously. This universality is remarkable, but it applies to spherically averaged profiles: averaging discards angular information and makes localised substructure less prominent. Moreover, the residual departures from NFW are neither random nor negligible; they contain information about the internal structure and recent growth of haloes.

One important departure from NFW concerns the curvature of the profile within the virialised region. Higher-resolution simulations showed that the logarithmic density slope continues to vary towards the smallest resolved radii rather than settling onto a well-defined inner power law \citep{navarro2004inner}. A profile in which the logarithmic slope itself varies as a power of radius provides a better description of this curvature. This is the Einasto law, originally introduced for stellar systems \citep{Einasto1965}; its additional shape parameter $\alpha$ controls how rapidly the slope changes with radius. \citet{Merritt2006} showed that the Einasto law provides a better overall description of simulated halo profiles than NFW and found evidence that more massive haloes have more strongly curved density profiles. The Einasto shape parameter therefore turns strict NFW universality into a low-dimensional family of halo profiles.

A distinct departure appears in the halo outskirts. Newly accreted particles pass through the halo and then reach the first apocentre of their orbit. As they slow near this turning point, they accumulate over a narrow radial range, producing a sharp steepening in the spherically averaged density profile. The corresponding splashback radius provides a dynamically motivated boundary between material that has orbited within the halo and material that is still falling in, and its location depends on the recent mass-accretion rate \citep{DiemerKravtsov2014,Adhikari2014,More2015}. A complementary result came from \citet{LucieSmith2022BuildingBlocks}. They trained a neural network to take the full three-dimensional matter distribution around each simulated halo and reproduce its spherically averaged density profile. The network needed only two latent degrees of freedom to describe profiles inside the virial radius, but a third for the outskirts; this additional degree of freedom was associated with infalling material. Halo density profiles are therefore low dimensional, but their remaining variations preserve information about halo growth. Yet even a precise density law describes only where the matter is, not how it moves. A dynamical description requires the distribution function (DF). Simple analytic forms for the DF of
simulated haloes have been fitted empirically \citep{HanyuHabe2001,Wojtak2008}
or derived from statistical mechanics and tested against simulations
\citep{HjorthWilliams2010}. Isotropic DFs measured from simulations have been
compared with NFW and DARKexp predictions and found to follow approximate
scaling relations \citep{Gross2024,Gross2025}, but no comparably simple DF has
been inferred from cosmological haloes without assuming its form. Finding such a form would extend the empirical simplicity of halo density profiles into phase space.

For a relaxed collisionless system, the DF specifies the full distribution of matter in position and velocity and therefore provides a complete dynamical description. All macroscopic properties follow from it: integration over velocity gives the spatial density, while velocity moments give the mean motion, velocity-dispersion tensor, and anisotropy at every position. For a steady, spherical, isotropic system, Jeans' theorem reduces the DF to a function $f(\varepsilon)$ of the relative binding energy, and Eddington inversion connects this function uniquely to a self-consistent density--potential pair \citep{Eddington1916,BinneyTremaine2008}. The spherical-isotropic approximation therefore turns a difficult phase-space problem into a one-dimensional one while retaining information that is absent from $\rho(r)$ alone.

There are indications that the regularity of DMO haloes extends beyond configuration space. In simulations, the pseudo-phase-space density $Q(r)=\rho(r)/\sigma^3(r)$ is close to a power law over a wide radial range \citep{TaylorNavarro2001}. Although $Q$ is not itself a DF, its simplicity motivates a search for similarly economical structure in $f$. Exact analytic DFs are known for special potential--density pairs, such as the Hernquist model, while DFs for cuspy halo profiles can be calculated by inversion and represented by numerical fitting formulae \citep{AnalyticalModelHernquist,Widrow2000}. Even in these idealised spherical systems, however, the DF is generally much less compact than the density law. A simple empirical expression for $f$ could therefore reveal phase-space structure that remains obscure in a numerical inversion or a high-dimensional interpolant.

This motivates the central question of this paper: after their dimensional scales have been removed, do approximately relaxed, Milky Way-mass DMO haloes share a low-complexity energy-space DF? Cosmological snapshots make this question non-trivial. Halo boundaries are ambiguous, weakly bound material is mixed with particles on first infall or fly-by trajectories, and individual haloes depart from exact equilibrium, sphericity, and isotropy. The inferred DF may consequently depend on how the bound component is defined, especially near the escape energy. One must therefore select a dynamically meaningful particle population, reconstruct its self-consistent potential and DF, and establish the degree of similarity between haloes before searching for a common analytic form.

We address this problem using 30 Milky Way-mass haloes from the DMO Auriga suite \citep{Grand2017Auriga}. We compare four particle-selection schemes, extract empirical DFs using the differential energy distribution and density of states, and rescale the haloes using their intrinsic density-profile scales. We then use PySR symbolic regression \citep{Cranmer2023PySR} to search the median similarity-scaled DF for expressions that balance accuracy against complexity. Symbolic regression has previously been applied to halo density profiles
\citep{Martin2026}, but not, to our knowledge, to the DF. Changes of energy variable guide the search towards the physical asymptotes, after which we retain only the common expressions that are well defined, monotonic and continuous towards the escape energy and that pass a loss cut. The remaining functional forms belong to simple extensions of the function form $f\propto\hat \varepsilon^{\nu}/(1-\hat \varepsilon)^{\mu}$. 

The paper is organised as follows. Section~2 defines the density profiles, relative energy, Eddington inversion, and density-of-states formalism used throughout, and introduces an analytic DF benchmark. Section~3 describes the Auriga simulations, the selection of bound particles, the extraction of empirical DFs, and the similarity scaling. Section~4 presents the physical admissibility conditions, symbolic-regression procedure, and resulting analytic expressions. Section~5 examines the density and velocity-dispersion profiles implied by these expressions, discusses the residual halo-to-halo variation, and summarises our conclusions.

\section{Distribution Functions}
We begin by defining the configuration-space profiles used to scale and assess the simulated haloes. The NFW and Einasto profiles are, respectively,
\begin{align}
\rho_{\mathrm{NFW}}(r) &=
\frac{\rho_s}{(r/r_s)(1+r/r_s)^2},
\label{eq:nfw}\\
\rho_{\mathrm{Einasto}}(r) &=
\rho_s \exp\!\left\{
-\frac{2}{\alpha}\left[\left(\frac{r}{r_s}\right)^\alpha - 1\right]
\right\},
\label{eq:einasto}
\end{align}
where $r_s$ and $\rho_s$ are the scale radius and scale density, with $d\ln\rho(r_s)/d\ln r=-2$, and $\alpha$ controls the shape of the Einasto profile. These density laws provide compact configuration-space descriptions, but the dynamical state of a collisionless system is specified by its DF $f(\mathbf{x},\mathbf{v},t)$, defined so that $f\,d^3\mathbf{x}\,d^3\mathbf{v}$ is the mass in a phase-space volume element. Its velocity moments determine the density, mean motion, velocity-dispersion tensor, and derived quantities such as the velocity anisotropy at every position. Together with Poisson's equation and the collisionless Boltzmann equation, these standard relations provide a closed description of a self-gravitating collisionless system; see \citet{BinneyTremaine2008}.

In this work, we restrict the general phase-space description to steady, spherically symmetric systems with isotropic velocities. In this limit, $\langle\mathbf{v}\rangle=0$, $\sigma_r^2=\sigma_\theta^2=\sigma_\phi^2$, and hence $\beta=0$. Jeans' theorem then reduces the DF to a function of a single variable, which we take to be the relative energy
\begin{equation}
\varepsilon = \Psi - \tfrac{1}{2}v^{2},
\end{equation}
where $\Psi=-\Phi+\Phi_0$ is the relative potential. For a self-consistent spherical-isotropic system, Eddington inversion gives \citep{Eddington1916}
\begin{equation}
\label{eddinversion}
f(\varepsilon) = \frac{1}{\sqrt{8}\pi^{2}}\frac{d}{d\varepsilon}\int_{0}^{\varepsilon}\frac{d\Psi}{\sqrt{\varepsilon - \Psi}}\frac{d\rho}{d\Psi}
\end{equation}
and hence determines $f(\varepsilon)$ when $\rho(\Psi)$ is known and the boundary conditions are specified. Directly differentiating an empirical density profile makes equation~\eqref{eddinversion} sensitive to numerical noise. We instead use the density of states
\begin{equation}
\label{densstates}
g(\varepsilon) = 16\pi^{2}\int_{0}^{r_{m}(\varepsilon)}dr\, r^{2}\sqrt{2[\Psi(r)-\varepsilon]},
\end{equation}
where $r_m(\varepsilon)$ is the largest radius accessible at relative energy $\varepsilon$. The differential energy distribution then gives
\begin{equation}
\label{energydiff}
dM/d\varepsilon = g(\varepsilon)f(\varepsilon).
\end{equation}

Equation~\eqref{energydiff} defines a distribution function for any steady spherical system, isotropic or not. For an anisotropic system it is the phase-space density averaged over each energy shell. Thus, it preserves $dM/d\varepsilon$ and the total mass exactly, reproduces $\rho(r)$ only approximately, and carries no information on velocity anisotropy. A useful closed-form benchmark is the Hernquist model \citep{AnalyticalModelHernquist}, whose density profile is
\begin{equation}
\rho_{\text{H}}(r) = \frac{M}{2\pi a^{3}}\frac{1}{(r/a)(1+r/a)^{3}}
\end{equation}
and whose isotropic DF is
\begin{multline}
\label{Hern}
f_{H}(\hat \varepsilon) = \frac{1}{\sqrt{2}(2\pi)^{3}\sqrt{G^{3}Ma^{3}}}\frac{\sqrt{\hat \varepsilon}}{(1- \hat \varepsilon)^{2}}\biggl[ (1-2\hat \varepsilon)(8\hat \varepsilon^{2} - 8\hat \varepsilon-3)\\
+\frac{3\arcsin(\sqrt{\hat \varepsilon})}{\sqrt{\hat \varepsilon(1 - \hat \varepsilon)}}\biggr]
\end{multline}
where \(\hat \varepsilon = \varepsilon/\Psi(0)\), with \(\Psi(0) = GM/a\), denotes the dimensionless relative energy. Even this deliberately simple density law produces a comparatively involved DF. For NFW-like cusps, the spherical DF can be obtained by inversion
\citep{Widrow2000,Lo01}, but its form is again substantially more complicated than the density profile. Its limits are nevertheless simple: a cusp $\rho\propto r^{-\gamma}$ gives $f\propto(1-\hat\varepsilon)^{-(6-\gamma)/(4-2\gamma)}$
as $\hat\varepsilon\to1$ \citep{Widrow2000}, and an envelope $\rho\propto\Psi^n$ gives $f\propto\hat\varepsilon^{\,n-3/2}$ as $\hat\varepsilon\to0$ (\citealt{BinneyTremaine2008}; see \citealt{Ev06}
for the anisotropic case).
\citet{Widrow2000} fitted NFW-like DFs with a template
$\hat\varepsilon^{3/2}(1-\hat\varepsilon)^{-\lambda}$ times slowly varying
corrections. This contrast motivates the empirical search developed below. Namely, rather than presupposing a density law and inverting it, we ask whether cosmological haloes themselves select a simpler common form in energy space.

\section{Constructing Empirical Distribution Functions}

We now construct numerical, empirical DFs directly from the simulated particle data. We first describe the Auriga DMO simulations and then compare alternative definitions of the bound halo population. For each definition, we reconstruct $\varepsilon \mapsto f(\varepsilon)$ without assuming an analytic form; these empirical DFs subsequently provide the target data for the symbolic regression in Section~4.

\subsection{Auriga DMO Simulations}
We analyse the 30 Milky Way-mass haloes in the dark-matter-only counterparts of the original Auriga suite at resolution level 4. Auriga comprises cosmological zoom-in simulations of isolated haloes selected at $z=0$ and evolved in a $\Lambda$CDM cosmology with the moving-mesh code \textsc{arepo} \citep{Grand2017Auriga}. We use the snapshots and group catalogues from the Auriga public data release \citep{Grand2024Auriga,AurigaWebsite}. Each halo contains approximately $(1$--$2)\times10^{7}$ dark-matter particles and has a mass of $(1$--$2)\times10^{12}\,M_{\odot}$. Unless stated otherwise, we identify the virial boundary with $R_{200,m}$, the radius within which the mean enclosed density is 200 times the cosmic mean matter density, and write $R_{\rm vir}\equiv R_{200,m}$ and $M_{\rm vir}\equiv M_{200,m}$.

\begin{figure*}
    \centering
    \includegraphics[width=\textwidth]{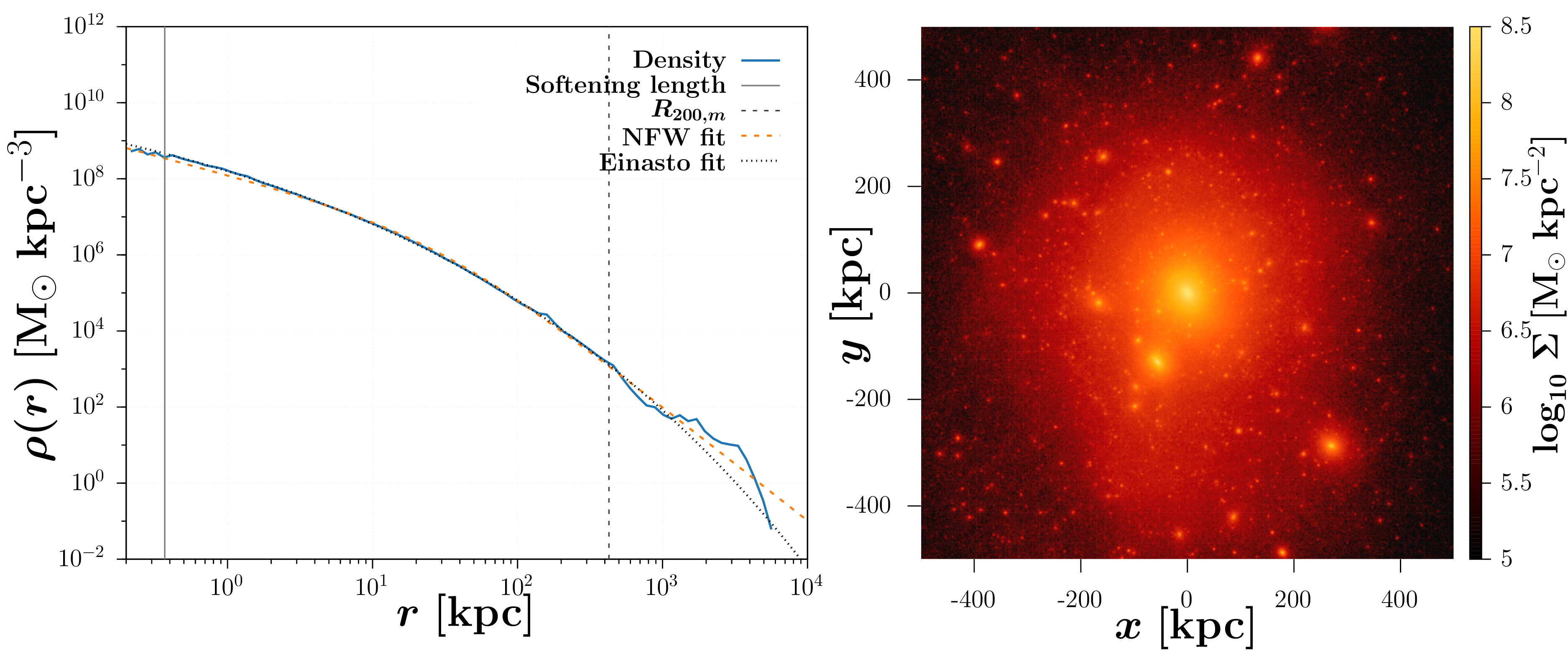}
    \caption{The density structure of Halo 02 in the Auriga DMO level-4 suite. (Left panel) \textit{Solid blue line}: average radial density profile computed in $80$ logarithmic bins for visualisation only. \textit{Dashed orange line}: NFW fit. \textit{Dotted dark line}: Einasto fit. \textit{Black vertical line}: simulation softening length, $\epsilon = 0.369\,{\rm kpc}$. \textit{Black dashed line}: virial radius, $R_{200,m} = 427\,{\rm kpc}$. (Right panel) Projected density along the $z$-axis within a square of side length $1000\,{\rm kpc}$.}
    \label{fig:halo02}
\end{figure*}

The Auriga group catalogues identify parent haloes with a friends-of-friends algorithm and gravitationally self-bound substructure with \textsc{subfind} \citep{Springel2001Subfind,Grand2024Auriga}. We centre each halo on the first, dominant entry in the catalogue's \texttt{Subhalo} group, using its \texttt{SubhaloPos} field, which records the position of the minimum-potential particle. Halo 02 serves as our representative reference system; Fig.~\ref{fig:halo02} shows both its spherically averaged density profile and its projected two-dimensional density distribution.

\subsection{Numerical Construction and Particle Selection}

Although Jeans' theorem reduces the DF to one dimension under our assumptions, measuring an empirical DF from a cosmological simulation remains non-unique: both the halo boundary and the meaning of a bound particle require operational definitions. We therefore compare several particle selections before measuring $f(\varepsilon)$ numerically. This procedure makes no assumption about the analytic shape of the DF; the symbolic representation is inferred only later from the measured curves.

The central regions of these haloes have relaxed through hierarchical assembly over a Hubble time, but many particles in their outskirts are still infalling or following fly-by trajectories. Because the theory on which we rely assumes a collisionless system in dynamical equilibrium, we seek a well-defined, self-consistent subset of particles that is gravitationally bound and approximately relaxed. This restriction approximates the equilibrium population while leaving the functional form of its DF unconstrained. Since the filtering introduces modelling assumptions, we compare four selection procedures.

The first is a \textit{Naive Radial Selection}, which retains every particle within the virial radius of the central subhalo, as defined by the group catalogue. The second is a \textit{Static Fixed-Point Selection}, which iteratively updates the pair \((\rho, \Psi)\) and retains bound particles, 
\(\varepsilon > 0\), with respect to the potential generated by the retained subset itself. The third and fourth are \textit{dynamic selections}, retaining particles whose trajectories show a pericentric passage followed by an apocentric passage: \textit{Dynamic Selection: Without Hubble Flow}, classified from peculiar radial velocities, and \textit{Dynamic Selection: With Hubble Flow}, classified from proper radial velocities including the Hubble flow. All trajectories are traced in the frame of the central subhalo through 51 N-body snapshots spanning $z = 1$ to $z=0$.

Once a bound particle subset is defined, we reconstruct its density profile and gravitational potential to obtain the empirical DF. We compute the radial density profile in 320 logarithmically spaced spherical shells, spanning the  $10^{-6}$ quantile of the particle-radius distribution to the outermost bound particle, so as to capture the dynamics of the halo from its innermost to outermost regions. This profile is then propagated into the empirical distribution function, whose inner and outer tails are correspondingly sensitive to these bounds. At the inner edge, we find the tail converges by the $10^{-4}$ quantile, on the order of the softening length; we retain the more conservative $10^{-6}$th quantile as a margin against this. At the outer edge, varying the bound instead flattens the tail of the distribution function. We fix this outer bound by our notion of a \textit{bound} particle, which we discuss below; any alternative outer bound instead reduces to a fixed radial cut, which we treat separately as the \textit{radial selection} method introduced above. We therefore accept the intrinsic noise contributed by outskirts material in order to construct a distribution function that remains robust out to these outskirts. 

Defining each bin centre as the geometric mean of its shell edges gives

\begin{equation}
\rho(r) = \frac{3\,M_{\mathrm{shell}}(r)}{4\pi \left(r_{\mathrm{out}}^{3} - r_{\mathrm{in}}^{3}\right)},
\end{equation}
where $r = \sqrt{r_{\mathrm{in}}\,r_{\mathrm{out}}}$, $r_{\rm in}$ and $r_{\rm out}$ are the shell edges, and $M_{\rm shell}$ is the mass in the shell. We then estimate the mass enclosed within $r$ as

\begin{equation}
M_{\text{enc}}(r) = \int_{0}^{r}  ds \cdot 4\pi s^{2}\rho(s)
\end{equation}
and obtain the relative potential under spherical symmetry from
\begin{equation}
\Psi(r) = -\Phi(r) = G\int_{r}^{R_{\text{max}}}ds \cdot \frac{M_{\text{enc}}(s)}{s^{2}} + \frac{GM_{\text{enc}}(R_{\text{max}})}{R_{\text{max}}}
\end{equation}
Here $R_{\max}$ is the distance from the central-subhalo centre to the outermost particle classified as bound. Above, we chose the gauge $\Phi(\infty) = 0$ which implicitly assumes that the density of the selected population vanishes beyond $R_{\max}$. The particle energies directly yield the cumulative and differential energy distributions, $M(\leq\varepsilon)$ and $dM/d\varepsilon$. For the DF calculation, velocities are peculiar velocities in the halo rest frame: we subtract the bulk velocity of the central subhalo and do not add the cosmological recession term $H(z)r$. The Hubble term enters only when defining the alternative proper-velocity dynamic selection above. 

Equations~\eqref{densstates} and \eqref{energydiff} then give the density of states and the empirical DF.  Beyond the truncation radius $R_{\max}$ we take $\rho=0$, so the potential there is that of a point mass, $\Psi(r)=GM_{\text{enc}}(R_{\max})/r$. Orbits with $\varepsilon<\Psi(R_{\max})$ therefore reach $r_m>R_{\max}$, as defined implicitly in \eqref{densstates}, and the vacuum part of the integral is evaluated analytically. We will however take $\max_{\mathrm{halo}} \bigl[\Psi(R_{\mathrm{max}})/\Psi(0)\bigr] \approx 0.03$ as a lower bound for the region over which we compute the distribution function.

Orbits with $\hat\varepsilon \lesssim \Psi(R_{\rm vir})/\Psi(0) \approx 0.14$ reach beyond $R_{\rm vir}$ and have radial periods comparable to or longer than the $\approx 8$ Gyr between $z=1$ and $z=0$. This weakly bound part of the DF is therefore not phase mixed. Its shape partly reflects recent accretion and the time window of the dynamic selection rather than an equilibrium population, and we treat constraints from this region with corresponding caution. We express energy as \(\hat\varepsilon=\varepsilon/\Psi(0)\). Because the logarithmic radial grid does not reach \(r=0\), we approximate $\Psi(0)$ by the potential at the innermost grid point, $r_{\min}\simeq30\,{\rm pc}$. This radius is about a factor of 12 smaller than the level-4 gravitational softening length, $0.369\,{\rm kpc}$; it is therefore a numerical proxy for the central potential, not a physically resolved radius.
\begin{figure}
\includegraphics[width=\columnwidth]{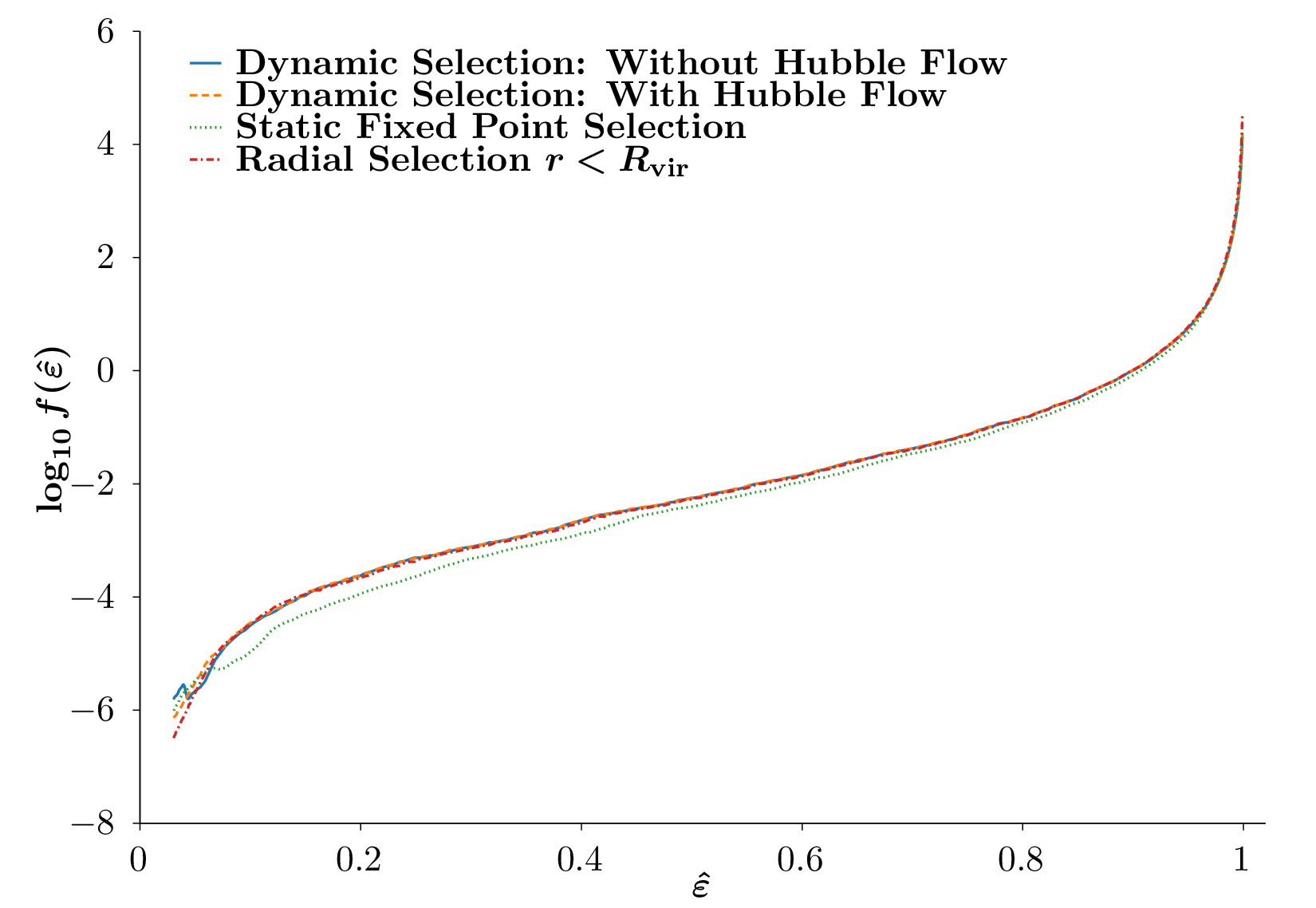}
    \caption{The empirical distribution function of Halo 02 as computed under different selection procedures.   \textit{Blue Solid Line}: Computed with dynamical selection using peculiar velocities. \textit{Orange Dashed Line}: Computed with dynamical  selection using proper velocities. \textit{Dotted Green Line}: Computed using the iterative fixed-point selection. \textit{Dashed-Dotted Red Line}: Naive Radial Selection.}
    \label{fig:df_selection_methods}
\end{figure}

Figure~\ref{fig:df_selection_methods} compares the four empirical DFs for Halo 02. The two dynamic selections and the naive radial selection agree to within approximately $0.05$\,dex over the region $0.07\lesssim\hat\varepsilon<1$, including the steep high-relative-energy tail, associated with the inner-halo region. They separate only in the weakly bound regime. The radial selection falls $0.2$--$0.4$\, dex below the dynamic selections for $\hat\varepsilon\lesssim0.05$. States with $\hat\varepsilon<\Psi(R_{\rm vir})/\Psi(0) \approx 0.14$ have turning points beyond $R_{\rm vir}$, so a hard radial boundary removes part of their population, while it can also admit
infalling or fly-by particles inside $R_{\text{vir}}$. This is the same limitation that motivates the splashback radius as a dynamically defined halo
boundary in the literature.

The fixed-point selection lies below the other selections, most strongly near $\hat\varepsilon\approx0.1$. Starting from all high-resolution particles, it retains about twice the bound mass, out to 8\,Mpc for Halo~02, so its potential
and density of states differ substantially from those of the other selections. The two dynamic selections yield near-identical DFs for $\hat\varepsilon\gtrsim0.1$, although their classifications disagree for about 15\% of all particles. Without the Hubble flow, the DF develops a kink at low relative energy, $\hat\varepsilon\approx0.04$. We interpret this feature as material in the outskirts of the central or neighbouring subhaloes whose peculiar motion mimics a pericentre and apocentre about the central halo even though the material is not orbiting it. Since particles at first apocentre also cluster near the splashback radius, we cannot fully exclude that this kink partly reflects genuine caustic structure rather than contamination alone; disentangling the two would require an independent splashback estimate for Halo 02, which we leave to future work. To avoid this bias, we adopt the dynamic selection based on proper radial velocities, including the Hubble flow, for the remainder of the analysis.
\subsection{Self Similarity}
The approximate self-similarity of DMO density profiles has been established over a wide range of halo masses \citep{Navarro1997NFW,Wang2020Halostructure}. We test whether this regularity extends from configuration space to the spherical, isotropic DF. Specifically, we compare two similarity transformations, one based on the virial radius (here $R_{200,c}$ from the group catalogue, with $\rho_*=\rho(R_{200,c})$) and one based on the intrinsic density-profile scale $r_s$, with each length paired with its corresponding density normalisation. For each halo, $(r_s,\rho_s)$ are the parameters of an NFW fit
(equation~\ref{eq:nfw}) to the density profile of its bound particles over $2\le r\le300$\,kpc, using shells with at least 200 particles and minimising the squared residuals in $\ln\rho$. Under the assumptions of spherical symmetry, velocity isotropy, and dynamical equilibrium, we hypothesise that the scaled DFs share a universal shape and test that hypothesis below.

Writing either pair of reference length and density scales as $(r_*,\rho_*)$, self-similarity implies that the DF at fixed dimensionless relative energy factorises into a universal shape function and a scale-dependent normalisation:
\begin{equation}
f(\hat \varepsilon; r_{*}, \rho_{*}) = A(r_{*}, \rho_{*})w(\hat \varepsilon)
\end{equation}
Dimensional analysis gives a global constant $C$ such that
\begin{equation}
A(r_{*}, \rho_{*}) = \frac{C}{G^{3/2}r_{*}^{3}\rho_{*}^{1/2}}
\end{equation}
Thus, we may transform the curves from the phase co-ordinates of halo $i$ to halo $j$ using,
\begin{equation}
f^{(j)}_{i}(\hat \varepsilon; r_{*,i},\rho_{*,i}) =\left(\frac{r_{*,i}}{r_{*,j}}\right)^{3}\left(\frac{\rho_{*,i}}{\rho_{*,j}}\right)^{1/2} f_{i}(\hat \varepsilon; r_{*,i},\rho_{*,i})
\end{equation}
The length scale should capture the halo structure relevant to the shared dynamics. We will consider our reference halo to be Halo 02, and thus define $f_{\rm sim}$ to be the pointwise median of the similarity transformed curves in the phase co-ordinates of Halo 02. An analytic regression of $f_{\rm sim}$ can therefore be mapped back to the DF of the $i$th halo through
\begin{equation}
\label{inverse_sim}
f_{i}(\hat \varepsilon; r_{*,i},\rho_{*,i}) = \left(\frac{r_{*,2}}{r_{*,i}}\right)^{3}\left(\frac{\rho_{*,2}}{\rho_{*,i}}\right)^{1/2} f_{\text{sim}}(\hat \varepsilon; r_{*,2},\rho_{*,2})
\end{equation}
Rewritten more formally, the above essentially claims
\begin{equation}
f_{i}(\hat \varepsilon; r_{*,i},\rho_{*,i}) = \frac{C}{G^{3/2}r_{*,i}^{3}\rho_{*,i}^{1/2}}w(\hat \varepsilon)
\end{equation}
where $w$ is the candidate universal shape function for spherical, isotropic DMO haloes.

As discussed in the previous section, we will construct distribution functions using bound particles classified with Hubble flow, on the interval $[0.03,0.999]$. The empirical curves are reliable over approximately $0.03\leq\hat\varepsilon\leq0.999$, thus we will also conduct the symbolic regression on this interval. All curves are linearly interpolated onto a common grid of 400 points uniformly spaced in $\hat\varepsilon$ over the range covered by every halo, $0.031 \le \hat\varepsilon \le 0.996$.

\begin{figure*}
    \centering
    \includegraphics[width=\textwidth]{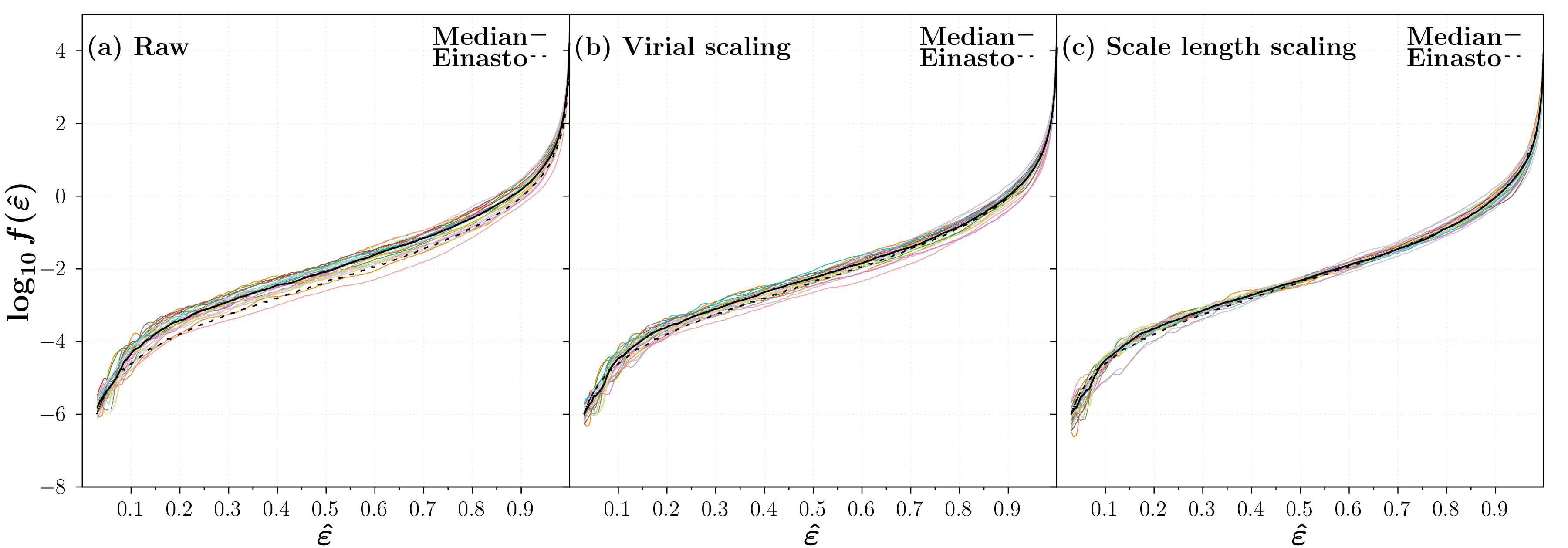}
    \caption{The empirical distribution functions of all 30 haloes under similarity transformations to the scales of Halo 02.
(a) The raw empirical distribution functions.
(b) $R_{200,c}$ scaling of the empirical distribution functions. 
(c) $(r_s,\rho_s)$ scaling of the empirical distribution functions. 
Coloured curves represent the individual haloes, while the \textit{Black Dashed Line} denotes the Einasto reference distribution function obtained from Eddington inversion of the fitted density profile of Halo 02. \textit{Black Solid Line}: The median of the distribution functions.}
    \label{fig:sim_scaling}
\end{figure*}

\begin{figure}
\includegraphics[width=\columnwidth]{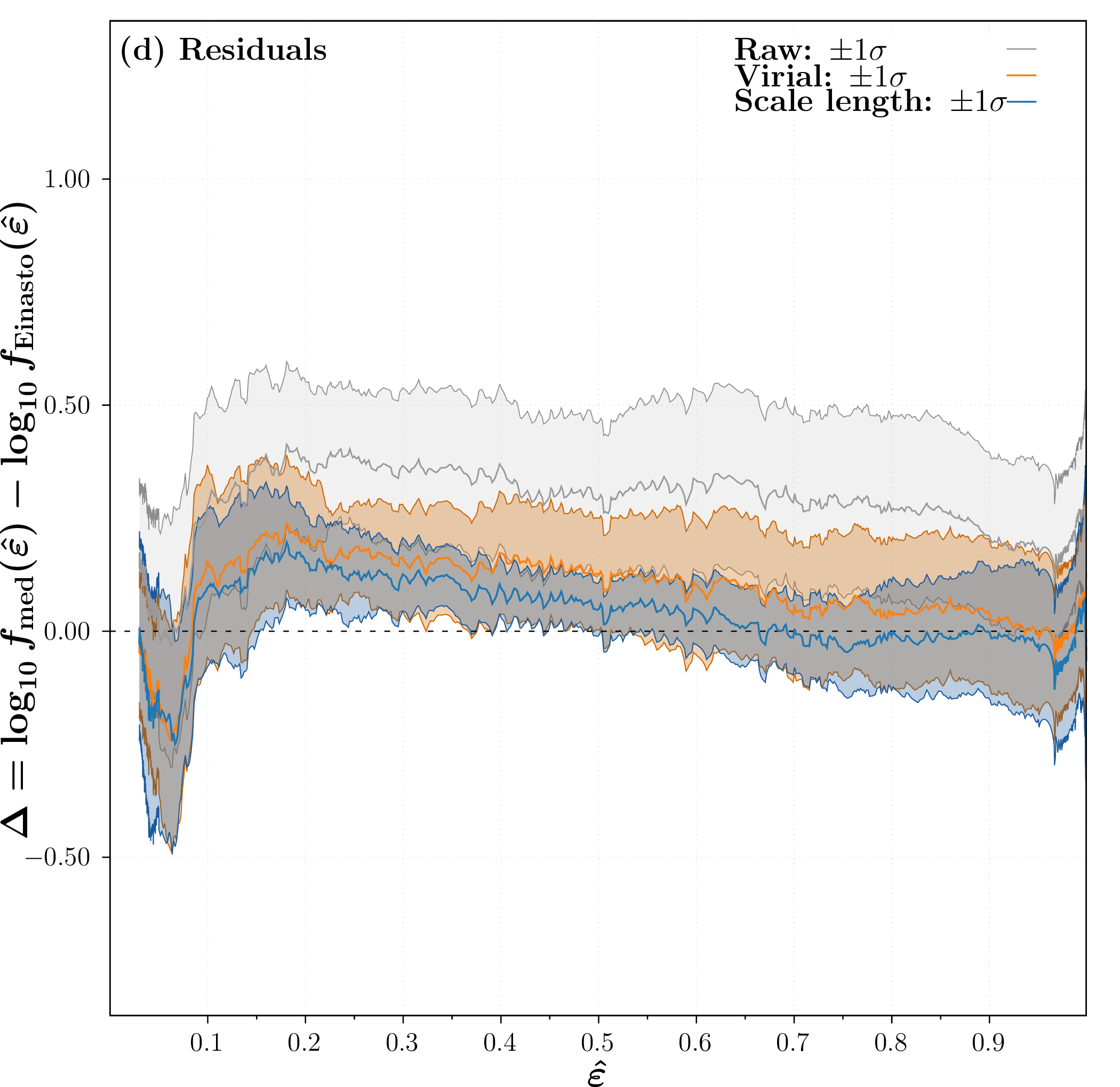}
    \caption{The residual of the similarity transformed curves, under the different choices of reference length, relative to the distribution function generated by an Einasto fit of Halo 02. Residuals relative to the Einasto reference, defined as $\Delta = \log f_{\mathrm{Med}} - \log f_{\mathrm{Einasto}}$. \textit{Grey Solid Line}: No scaling, region indicates $\pm1\sigma$. \textit{Blue Solid Line}: $r_{s}$ scaling, shaded region indicates $\pm1\sigma$. \textit{Orange Solid Line}: $R_{200,c}$ scaling, shaded region indicates $\pm1\sigma$.}
    \label{fig:residuals}
\end{figure}

Figures~\ref{fig:sim_scaling} and \ref{fig:residuals} show that
scaling by $(r_s,\rho_s)$ yields less halo-to-halo scatter and smaller
residuals relative to the Einasto reference. Figure~\ref{fig:sim_scaling} also shows, solely as a comparison baseline, the DF obtained by Eddington inversion of the best-fitting Einasto density profile for Halo 02. This Einasto curve is neither used to define the similarity transformation nor fitted by PySR. Before applying either similarity transformation, the DFs already
occupy a relatively narrow band. Although the Auriga haloes span a
narrow mass range, they still differ in virial radius and
concentration; the variation in concentration, in particular, reflects
their distinct assembly histories. Both similarity transformations
tighten the ensemble, but the $(r_s,\rho_s)$ transformation produces
the strongest collapse towards a common shape. This tighter collapse has
a natural interpretation. $R_{200,c}$ scaling removes differences in halo mass and overdensity radius but does not explicitly account for variations in concentration, whereas \((r_s,\rho_s)\) aligns the internal density structure of the haloes. Since the spherical-isotropic DF is fixed by the density–potential pair, a scaling that better aligns the density profiles should also better align the DFs. Exact scatter collapse is not expected because of finite particle sampling and because the haloes depart to varying degrees from sphericity, isotropy, and equilibrium. We therefore adopt the median of the \((r_s,\rho_s)\)-scaled DFs as the target for symbolic regression.

\section{Analytic distribution functions with symbolic regression}
Having identified a common shape for the similarity-scaled DFs, we now
seek a compact analytic representation. Our regression
target, \(f_{\rm sim}\), is the median of the \((r_s,\rho_s)\)-scaled DFs expressed in the characteristic scales of Halo 02. We choose Halo 02 as the reference because its empirical DF is the smoothest in the sample and the most stable across the particle-selection methods considered above. With the aim of finding simple functional forms that hold out of sample, we conduct a 6-fold cross validation procedure. Namely, we split the 30 haloes into 6 folds of 5, and in each fold train on the remaining 25 haloes and validate on the 5 held out, all in Halo 02's phase co-ordinates; the regression target in each fold is the median of its 25 training haloes. On the sets of training haloes we perform symbolic regression using PySR, a recently developed genetic symbolic regression algorithm, written in Julia with a Python frontend, which we detail in Appendix A. We regress $\log_{10} f$ rather than $f$, since $f$ spans many orders of magnitude.

\subsection{Physicality}

PySR is exceptionally powerful at discovering compact analytic
expressions within a vast search space, but it does not automatically
enforce the physical behaviour we require of a candidate DF. Numerical accuracy
alone therefore does not guarantee that a candidate expression is
physically admissible: it may have inappropriate behaviour near an
energy boundary, fail to be real and finite within the physical
domain, or offer no simple stability guarantee. We assess the
candidate expressions against three constraints: $f(\hat\varepsilon)
\to 0$ as $\hat\varepsilon\to 0$, strict monotonicity, and
well-definedness throughout $0<\hat\varepsilon<1$.

The first constraint governs the weak-binding limit. Requiring the DF
to approach zero continuously as $\hat\varepsilon\to0$ excludes models
that remain finite up to the escape energy and must then be truncated
abruptly. This is a modelling choice rather than a necessary property
of every equilibrium DF, but it is appropriate for the idealised halo models that we seek to approximate. At the
opposite boundary, $\hat\varepsilon\to1$, we do not require the DF to
diverge, since doing so could build a central density cusp into the
search and exclude admissible cored models. Moreover, the endpoints
lie outside the energy range over which the empirical DFs are reliable.
The boundary constraints should therefore be understood as conditions
on the analytic continuation of the fitted DF, rather than as direct
measurements of the limiting behaviour of the finite simulated haloes.

The second constraint, $\mathrm{d}f/\mathrm{d}\hat\varepsilon>0$,
requires phase-space density to increase with binding energy. Because
our relative energy satisfies $\varepsilon=-E+\mathrm{const}$, this is
equivalent to $\mathrm{d}f/\mathrm{d}E<0$ in the conventional energy
variable. For a spherical, isotropic equilibrium, this monotonicity is
a sufficient, though not necessary, condition for linear stability
against radial and non-radial perturbations
\citep{Antonov1962,Doremus1971,DejongheMerritt1988}. We therefore use
it as a conservative stability criterion within the assumptions of
this work. Finally, well-definedness excludes expressions that become
complex, singular, or discontinuous anywhere inside the physical
energy domain. Table~\ref{tab:physicality_conditions} summarises these
three constraints.

\begin{table}
    \centering
    \begin{tabular}{@{}p{0.3\linewidth} p{0.3\linewidth} p{0.3\linewidth}@{}}
        \hline
        Condition & Requirement & Motivation \\
        \hline
        Low-energy behaviour
        & \(f(\hat{\varepsilon}) \to 0\) as \(\hat{\varepsilon} \to 0\)
        & Ensures a continuous weak-binding tail at the escape energy \\[0.5em]

        Monotonicity
        & \(\mathrm{d}f/\mathrm{d}\hat\varepsilon>0\) on \((0,1)\)
        & Provides a sufficient linear-stability condition in the spherical-isotropic limit \\[0.5em]

        Well-definedness
        & \(f(\hat{\varepsilon})\) is well defined for all \(\hat{\varepsilon}\in(0,1)\)
        & Excludes complex, singular, or discontinuous behaviour within the physical domain \\[0.5em]
        \hline
    \end{tabular}
    \caption{Physicality conditions imposed on candidate symbolic models.}
    \label{tab:physicality_conditions}
\end{table}
\subsection{Training}
In an attempt to address the obvious degeneracy of symbolic regression, we linearised the asymptotics at the boundaries by considering a family of changes of variables and regressed symbolically using this transformed variable, effectively a change of co-ordinates. The family of bijections $T: (0,1) \mapsto \mathbb{R}$ stretches the boundaries, amplifies local curvature and encourages PySR to search for \textit{physical} expressions. We used five such transformations, these were:
{\small
\[
\left\{\ln\left(\frac{\hat \varepsilon}{1-\hat \varepsilon}\right), \sqrt{\frac{\hat \varepsilon}{1-\hat \varepsilon}},   \tan\left[\pi\left(\hat \varepsilon-\frac{1}{2}\right)\right], 
\frac{\hat \varepsilon-\frac{1}{2}}{\hat \varepsilon\left(1-\hat \varepsilon \right)}, \ln\left[\frac{\ln\left(\hat \varepsilon\right)}{\ln\left(1 - \hat \varepsilon\right)}\right]\right\}
\]
}
In addition to these, we also regressed the median directly. In each PySR run, we regressed using the following family of unary operators  $\{ \log, \exp, \sqrt{\cdot}, \arcsin, \sinh \}$, with strict nesting constraints and low complexity exponents, see Appendix A for details on hyperparameters used. For each run, PySR selects expressions on the \textit{Pareto} front, meaning the set of models that are not outperformed simultaneously in both accuracy and complexity. This naturally optimises for both the complexity, measured as a function of the lengths and types of operators used in the expression, and the MSLE loss relative to the dataset regressed. From each run, after removing expressions of complexity above $20$ in the regressed co-ordinate and transforming back to the original variable, we retain expressions for which the low-energy limit is established symbolically and no violation of monotonicity or well-definedness is found on a dense sweep of $(0,1)$. The low-energy limit is established symbolically, while monotonicity and well-definedness are tested at approximately $80,000$ points spanning $10^{-300}\le\hat\varepsilon\le1-10^{-300}$ using SymPy and mpmath libraries. Note that the latter two conditions are necessary but not sufficient.  

Denoting by $\mathcal{L}$ the mean squared $\log_{10}$ error (MSLE), averaged
over the $K$ haloes of the relevant set and
the $N$ grid points, namely
\begin{equation}
  \mathcal{L} = \frac{1}{K}\sum_{k=1}^{K}\frac{1}{N}\sum_{j=1}^{N}
  \left[\log_{10} f_{\rm model}(\hat\varepsilon_j)
        - \log_{10} f_{k}(\hat\varepsilon_j)\right]^2 ,
  \label{eq:loss}
\end{equation}
where $f_k$ is the similarity-scaled DF of halo $k$, and $f_{\rm model}$ is
the model mapped into the same co-ordinates via equation \eqref{inverse_sim}. We denote the
complexity, $\mathcal{C}$, defined as PySR's node count. Then the expressions can be ordered by $\mathcal{C}_{i}$ and represented as a set of tuples $\{(\mathcal{L}_{i}, \mathcal{C}_{i})\}_{i}$. Hence, once physical expressions are collected within each of the $6$ cross validation folds, we construct a Pareto front of the physical models, in loss-complexity space, and retain functional families, on the Pareto fronts, that are common across each of the folds. This not only improves the robustness of these models but also prioritises simple and more interpretable functional forms. 

Finally, we refit the remaining functional forms to the median of all 30 haloes, with $\mathcal{L}$ now the MSLE against this median; parameter uncertainties are obtained by bootstrap resampling of the 30 haloes, quoting the larger of the deviations of the 16th and 84th percentiles. The expressions can then be ranked using PySR's score function
\begin{equation}
\label{score}
S = -\frac{\ln(\mathcal{L}_{i}/\mathcal{L}_{i-1})}{\mathcal{C}_{i} - \mathcal{C}_{i-1}}
\end{equation}
This provides a final ranking of expressions. Families whose refitted Av. Halo MSLE exceeds ten times the minimum are reported but not carried forward.
\subsection{Results}
As discussed in the method, we considered a family of bijections mapping from $(0,1)$ to $\mathbb{R}$, so as to facilitate PySR's symbolic regression search. We found that the logit-based transformations, as well as the median itself, most effectively unravelled the asymptotic properties of the dataset whilst also promoting functions with the correct physical asymptotes. In fact, both $\log$ based transformations generated quasi linear datasets in the transformed space whilst the square root transformation effectively removed the asymptote at 1. This is partly expected, since in the logit variable $\log f$ of this family is linear, although the family also emerges without any change of variable.

Within a given fold, we collect the \textit{Hall of Fame} tables, as PySR calls them, generated for each of the five regressions, as well as the regression of the median itself, and remove expressions of complexity greater than $20$ in the regressed variable. This forms a large collection of $\sim 80$ expressions, for a total of $\sim 500$ expressions across all folds. From the tables, we notice that the expressions achieving the highest score, computed internally by PySR, are indeed low complexity expressions. We notice further that most of the physical expressions are generated from the logit transformed dataset, as well as the median. Within these folds, the physical expressions are then collected, simplified and their complexity recomputed, using SymPy, into a single table of $\sim 30$ expressions, with $\log_{10}$-complexities spanning the interval $[1,54]$. Despite our best efforts to promote physical expressions using these transformations, $\sim 50\%$ were deemed physical, however the log-based transformations achieved a physicality rate of  $\sim 70\%$ justifying our use of change of variables. The majority of the highly complex expressions were unphysical despite achieving smaller losses. We argue that, despite these transformations, PySR overfit local curvature so as to achieve greater numerical precision but introduced awkward asymptotic properties in the functions or a finite limit in $\log_{10}$ space at $\hat \varepsilon = 0$. Furthermore, several functions were not well-defined on $[0,1]$ due to PySR introducing nested expressions. 

\begin{figure}
    \includegraphics[width=\columnwidth]{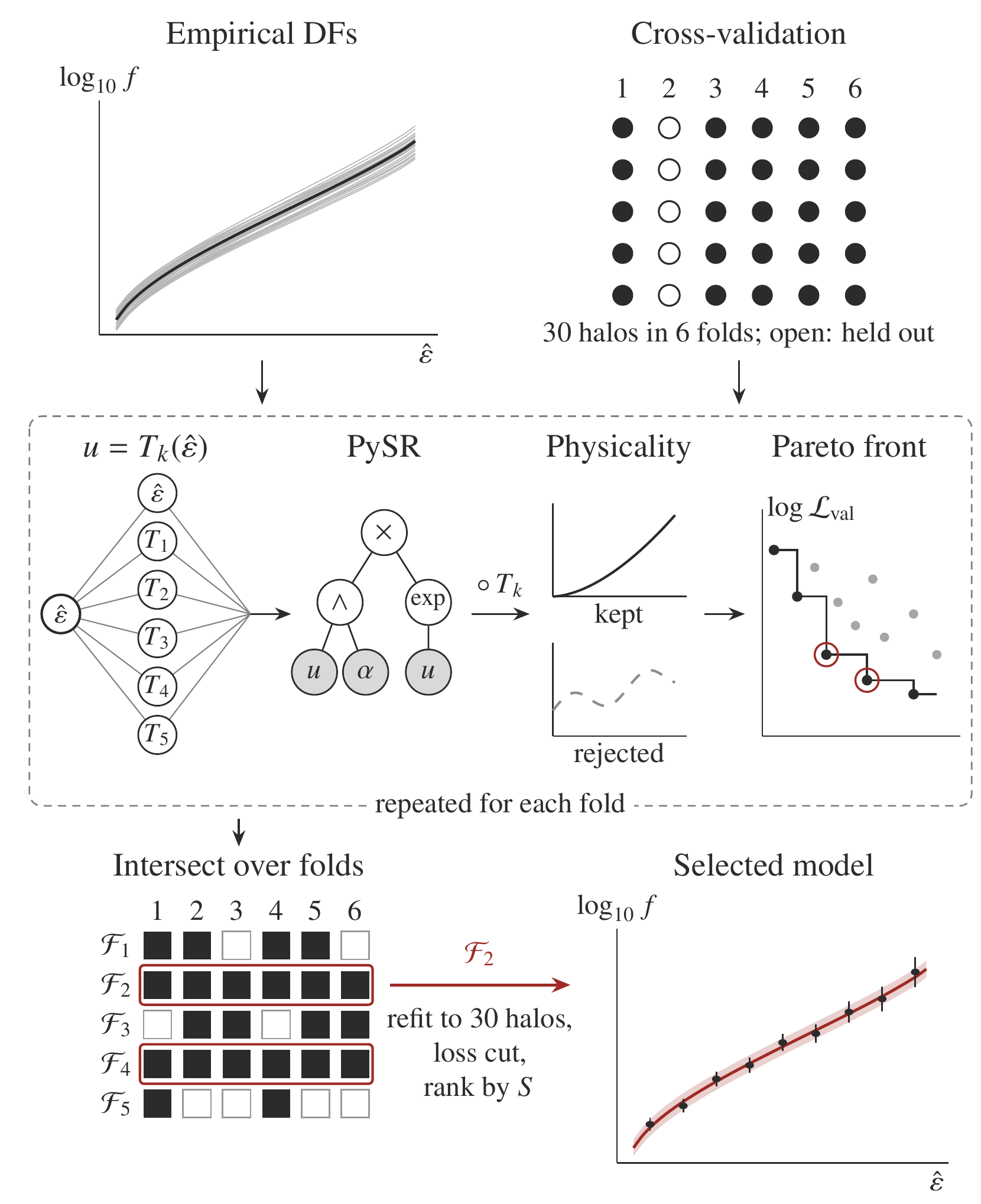}
    \caption{A general illustration of the training and testing pipeline with which we generate robust symbolic models.}
    \label{flowdiagram}
\end{figure}

Taking these expressions generated by PySR and evaluating them on the associated held out haloes of a given fold, Pareto fronts were formed for each fold. In every fold, the logit expression lowers the loss relative to the best power law with free normalisation by a factor of $64$–$74$ on the training median and $11$–$30$ on the held-out haloes. More complex front members improve the held-out loss by a further $16$–$40\%$. Summed over folds, $95\%$ of this gain comes from $\hat\varepsilon < 0.14$. The same logit family is found without any change of variable in two of the six folds. As discussed above, function forms common to each of the folds were retained and refit, with uncertainty in parameters, by bootstrapping the median of the entire dataset. Before moving on to the expressions found, Fig. \ref{flowdiagram} serves as a general illustration depicting the training process conducted in this work.

Refitting the common physical functional forms found on the Pareto fronts of each fold, and scoring using PySR's scoring function \eqref{score}, generates Table~\ref{winners}. The table illustrates the score and complexity of expressions, the latter again computed from the log-transformed expressions, as well as their average MSLE loss across the haloes. The final column illustrates the parameters associated to these models, as well as their uncertainty. The intersection of the fold Pareto fronts contains four physical families, listed in Table~\ref{winners} in order of complexity below the constant baseline. Only the two satisfying Av.\ Halo MSLE $\le 0.170$ are carried forward; the two simplest families fail this cut. Scores are computed from the MSLE against the 30-halo median, not the Av. Halo MSLE shown.  

\begin{table*}
    \centering
    \small
    \setlength{\tabcolsep}{6pt}
    \renewcommand{\arraystretch}{1.3}
    \begin{adjustbox}{max width=\textwidth}
    \begin{tabular}{l c c c l}
    \hline
    Expression & Complexity of $\log_{10} f$ & Score & Av.\ Halo MSLE & Parameters \\
    \hline
    $\displaystyle \frac{C}{G^{3/2}r_{s}^{3}\rho_{s}^{1/2}}$
    & 1 & -- & --
    & $C = 2.502\times 10^{-3}$ \\
    \hline
    $\displaystyle \frac{C}{G^{3/2}r_{s}^{3}\rho_{s}^{1/2}}\,
      \hat\varepsilon^{\ln 10}$
    & 2 & 0.029 & 2.658
    & $C = 0.391$ \\
    \hline
    $\displaystyle \frac{C}{G^{3/2}r_{s}^{3}\rho_{s}^{1/2}}\,
      \hat\varepsilon^{\nu}$
    & 4 & 0.849 & 0.483
    & $\displaystyle \begin{aligned}
      C &= 0.391\\
      \nu &= 5.132\pm0.019
    \end{aligned}$ \\
    \hline
    $\displaystyle \frac{C}{G^{3/2}r_{s}^{3}\rho_{s}^{1/2}}
      \left(\frac{\hat\varepsilon}{1-\hat\varepsilon}\right)^{\ln 10}$
    & 9 & 0.888 & 0.0214
    & $C = (1.920\pm0.071)\times 10^{-3}$ \\
    \hline
    $\displaystyle \frac{C}{G^{3/2}r_{s}^{3}\rho_{s}^{1/2}}
      \left(\frac{\hat\varepsilon}{1-\hat\varepsilon}\right)^{\ln 10}
      e^{-\hat\varepsilon_{0}/\hat\varepsilon}$
    & 13 & 0.219 & 0.0170
    & $\displaystyle \begin{aligned}
      C &= (2.134\pm0.092)\times 10^{-3}\\
      \hat\varepsilon_{0} &= 0.0290\pm0.0046
    \end{aligned}$ \\
    \hline
    \end{tabular}
    \end{adjustbox}
    \caption{The leaderboard of common functional expressions, found across Pareto fronts, trained and evaluated on the different cross-validation folds. The first row is a constant baseline, not a candidate, included so that every score is defined; scores are computed relative to the preceding row from the MSLE against the 30-halo median. The second and third rows have no free normalisation in $\log_{10} f$, so $C = 0.391$ is fixed by the units of $f$ rather than fitted. The baseline and the two simplest families fail the loss cut and are not generalised.}
    \label{winners}
\end{table*}
In light of these results, the remainder of this paper is devoted to analysing sources of error as well as the physical consequences of these analytic expressions and what they imply. So far, we have developed a pipeline, using PySR, that generates promising functional forms for Auriga DMO halo distribution functions. With these expressions in mind, we reconsider their general functional form, refit them on the bootstrapped median and investigate physical consequences for these expressions. Not only does this general functional form allow us to generalise the physical implications in the following section, but we avoid the non-physical consequences of our pipeline. For example, the exponent $\ln10$ in Table~\ref{winners} is an artefact of the complexity penalty rather than a physical constant. As we regress $\log_{10} f$, the expression $\log_{10} f = \ln(\hat \varepsilon/(1- \hat \varepsilon)) + \text{const}$ corresponds to $f \propto (\hat \varepsilon/(1- \hat \varepsilon))^{\ln 10}$, and PySR prefers the bare $\ln$ operator to a fitted coefficient multiplying $\log_{10}$, which would cost additional nodes. We therefore free the exponent and refit, giving the generalised forms in Table~\ref{tab:generalised}. We additionally include Form 2 as the natural generalisation of Form 1, decoupling its low- and high-energy exponents. The cutoff scale of Form 3, $\hat\varepsilon_0 \approx 0.03$--$0.04$, coincides with the lower edge of the fitted range. There, particles have completed less than one radial orbit since $z=1$. Its modest improvement in loss over Form 1 comes mainly from these lowest-energy bins, which are the least reliable part of the data. We therefore treat the factor $e^{-\hat\varepsilon_0/\hat\varepsilon}$ as an empirical description of the weakly bound tail rather than a \enquote{physical feature} of the DF. 

As a comparison for the values provided in Table~\ref{tab:generalised}, if we simply transform the numerical global median of the curves, in the coordinates of Halo 02, we obtain an average Halo MSLE of $0.0153$. Similarly, if one fits an Einasto profile onto the median of the distribution functions in Halo 02's co-ordinates, and similarity transforms its associated distribution function, we achieve an average Halo MSLE of $ 0.0196$. 

\begin{figure*}
    \centering
    \includegraphics[width=\textwidth]{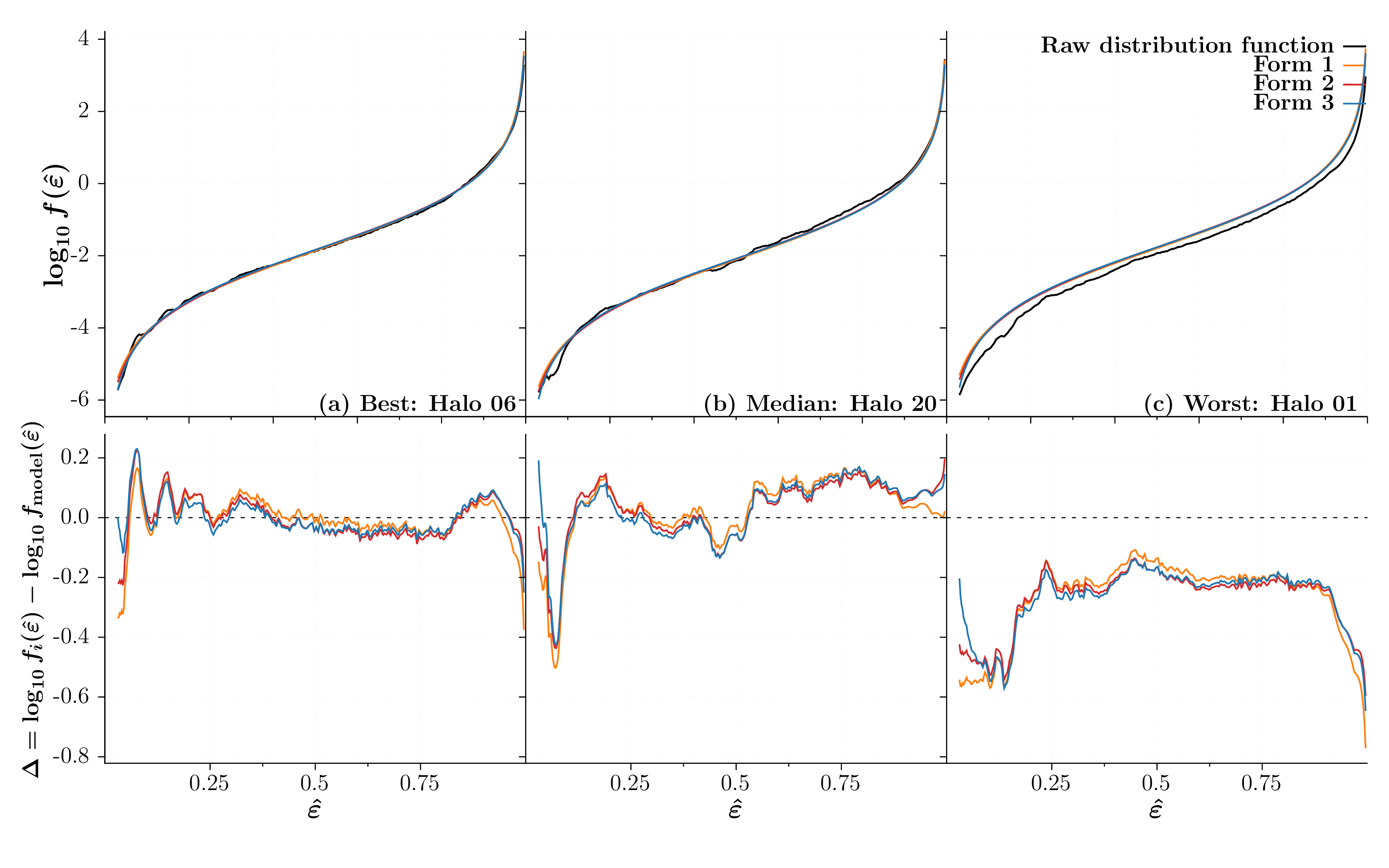}
    \caption{A plot of the best, median and worst symbolic fits, with respect to the leading model's MSLE. \textit{Solid Orange Line}: The plot of form 1 in $\log_{10}$ space as well as its residuals against the raw halo distribution function.  \textit{Solid Red Line}: The plot of form 2 in $\log_{10}$ space as well as its residuals against the raw distribution function. \textit{Solid Blue Line}: The plot of form 3 in $\log_{10}$ space as well as its residuals against the raw distribution function.}
    \label{MODELSPLOTTED}
\end{figure*}

Form~1 thus matches the Einasto DF at equal parameter count and improves on the one-parameter NFW
DF. The Hernquist DF does worse only because its exponents are
fixed at $5/2$ (Form~1 with $\nu=5/2$ gives an Av. Halo MSLE of 0.0300). Unlike the Einasto and NFW
DFs, which require numerical inversion, Form~1 and its implied $\rho(\Psi)$ and
$\sigma_r^2(\Psi)$ are closed form, and Forms~1--3 are analytically positive and
monotonic.  To visualise these results, we plot these generalised expressions found against a selection of raw distribution functions in $\log_{10}$ as shown in Fig. \ref{MODELSPLOTTED}. More specifically, to illustrate the robustness of these expressions, and hence consequently the accuracy of self similarity, we illustrate the best, median and worst fits achieved with respect to the MSLE of form 1. Since we have fit these models to maximise numerical accuracy, these models appear almost perfectly overlaid in the top panel of plots. Hence, it is necessary to provide residuals to distinguish the models' behaviour. Since all three models admit near-identical shapes on approximately
$[0.1, 0.95]$ and their residuals share the same peaks and troughs, these features belong to the individual empirical DFs. We note interestingly however these analytic expressions differ in the outer regions and generate tails at low $\hat \varepsilon$ of different steepness. We will discuss and analyse sources of such noise in the following section. 
\begin{table*}
    \centering
    \small
    \setlength{\tabcolsep}{6pt}
    \renewcommand{\arraystretch}{1.3}
    \begin{adjustbox}{max width=\textwidth}
    \begin{tabular}{c l c c l}
    \hline
    Form & Expression & Complexity of $\log_{10} f$ & Av.\ Halo MSLE &
  Parameters \\
    \hline
    1 &
    $\displaystyle \frac{C}{G^{3/2}r_{s}^{3}\rho_{s}^{1/2}}
      \left(\frac{\hat\varepsilon}{1-\hat\varepsilon}\right)^{\nu}$
    & 11 & 0.0200
    & $\displaystyle \begin{aligned}
    C &= (1.910\pm0.067)\times 10^{-3}\\
    \nu &= 2.347\pm0.019
    \end{aligned}$ \\
    \hline
    2 &
    $\displaystyle \frac{C}{G^{3/2}r_{s}^{3}\rho_{s}^{1/2}}
      \frac{\hat\varepsilon^{\nu}}{(1-\hat\varepsilon)^{\mu}}$
    & 13 & 0.0178
    & $\displaystyle \begin{aligned}
    C &= (2.47\pm0.12)\times 10^{-3}\\
    \nu &= 2.499\pm0.035\\
    \mu &= 2.226\pm0.048
    \end{aligned}$ \\
    \hline
    3 &
    $\displaystyle \frac{C}{G^{3/2}r_{s}^{3}\rho_{s}^{1/2}}
      \left(\frac{\hat\varepsilon}{1-\hat\varepsilon}\right)^{\nu}
      e^{-\hat\varepsilon_{0}/\hat\varepsilon}$
    & 15 & 0.0168
    & $\displaystyle \begin{aligned}
    C &= (2.193\pm0.058)\times 10^{-3}\\
    \nu &= 2.276\pm0.033\\
    \hat\varepsilon_{0} &= 0.0357\pm0.0081
    \end{aligned}$ \\
    \hline
    Einasto & N/A & N/A & 0.0196
    & $\displaystyle \begin{aligned}
    \alpha &= 0.1948\pm0.0066\\
    r_{s} &= 31.6~\mathrm{kpc}\\
    \rho_{s} &= (7.60\pm0.57)\times 10^{5}~\mathrm{M_\odot\,kpc^{-3}}
    \end{aligned}$ \\
    \hline
    NFW & N/A & N/A & 0.0223
    & $\displaystyle \begin{aligned}
    r_s &= 27.1~\mathrm{kpc}\\
    \rho_s &= (3.18\pm0.24)\times 10^{6}~\mathrm{M_\odot\,kpc^{-3}}
    \end{aligned}$ \\
    \hline
    Hernquist & \eqref{Hern} & 41 & 0.0304
    & $\displaystyle \begin{aligned}
    a &= 66.5~\mathrm{kpc}\\
    M &= (2.27\pm0.17)\times 10^{12}~\mathrm{M_\odot}
    \end{aligned}$ \\
    \hline
    \end{tabular}
    \end{adjustbox}
    \caption{Generalised analytic DFs refitted to the median of all 30
$(r_s,\rho_s)$-scaled haloes. The last three rows fit the same median with the
isotropic Einasto, NFW and Hernquist DFs, radii fixed at Halo~02's density-fit
values, each in its own $\hat\varepsilon=\varepsilon/\Psi(0)$, with only the
normalisation (and $\alpha$ for Einasto) free.}
    \label{tab:generalised}
  \end{table*}

\section{Discussion \& Conclusions}

\subsection{Physical Interpretations}
Returning to the original question at hand, we ask whether these symbolic fits in phase space yield interpretable functional forms for DMO distribution functions and where the implied physical observables, such as the radial density profile, agree with known results. The distribution function is known to satisfy the following pair of equations
\begin{gather}
\rho(\Psi) = 4\sqrt{2}\pi \int_{0}^{\Psi}d\varepsilon\, f(\varepsilon)\sqrt{\Psi - \varepsilon}, \label{rhopsi}\\
\nabla^{2}\Psi = -4\pi G\rho .
\end{gather}
The analytic expressions of Table~\ref{tab:generalised} are centred around the functional family of the form \(\hat \varepsilon^{\nu}/(1-\hat \varepsilon)^{\mu}\), with form 3 admitting an additional factor. This analytic form will be the focus of this section.

To solve this pair of equations, it is useful to work with the dimensionless
  variables \(\hat r = r/r_{s}\), \(\hat \rho = \rho/\rho_{s}\),
  \(\hat\varepsilon = \varepsilon/\Psi(0)\) and \(\hat \Psi = \Psi/\Psi(0)\), so
  that \(\hat\Psi(0)=1\) and the distribution function keeps the argument in
  which it was fitted. The structural energy scale \(G r_{s}^{2}\rho_{s}\) and
  the central potential differ by $q:= \Psi(0)/G r_{s}^{2}\rho_{s}$ which is not unity in general. Across the 30 haloes, $q$ has median $12.10$, ranging within $[11.75,12.81]$
and $\sigma_{q}/\langle q\rangle = 0.017$, so normalising energies
by $\Psi(0)$ is equivalent to the structural
scale, up to this accuracy. An untruncated NFW profile has \(q = 4\pi\), whilst Halo 02 has \(q = 11.95\). We adopt this value throughout when solving Poisson’s equation below and note that the two largest values belong to Haloes 01 and 18,
the two worst-fitting haloes, which we come back to in Section~\ref{sec:residual}. 

 Then, for the general distribution function
  \(f(\hat \varepsilon) \propto \hat
\varepsilon^{\nu}/(1-\hat
  \varepsilon)^{\mu}\), we find the density--potential relation satisfies
  \begin{multline}
  \label{denspotdf}                     
  \hat\rho(\hat \Psi) = 4\sqrt{2}\pi C q^{3/2}\hat
  \Psi^{\nu+3/2}\\\cdot\mathcal{B}(\nu+1,
  3/2){}_2F_{1}(\nu+1,\mu;\nu+5/2;\hat \Psi)
  \end{multline}
  where ${}_2F_{1}$ denotes the hypergeometric function, and Poisson's equation
  becomes \(\hat\nabla^{2}\hat\Psi = -(4\pi/q)\,\hat\rho\). We finally use a
  numerical scheme detailed in Appendix C to solve Poisson's equation under
  spherical symmetry with two initial constraints on $\hat \Psi$ at the cusp derived from \eqref{denspotdf}. We note first that the inner
  asymptotic slopes of the implied radial density, $\gamma_{\text{in}}$, can be shown to satisfy
\begin{equation}
\gamma_{\text{in}} = 2\left(\frac{\mu - 3/2}{\mu - 1/2}\right)
\end{equation}
for the general function form given above. The outer density slope, however, depends on which branch of Poisson's equation the singular-cusp solution follows. These include a \textit{truncated solution}, a \textit{Keplerian solution} and a \textit{scale-free solution}. We will discuss these, and their associated outer slopes, in more detail in Appendix B, however we do not consider the scale-free solution further, since it has
infinite mass and is not reached by the cusp solution. On the Keplerian
branch, Forms 1 and 2 have finite mass for $\nu>3/2$, which is satisfied. In the Keplerian case, 
\begin{equation}
\gamma_{\text{out}} = \nu +3/2
\end{equation}
Because Form 1 lies within $1\sigma$ of the critical exponent $\nu_c$ (Appendix B), the Keplerian branch is not generic, and we do not tabulate $\gamma_{\rm out}$. Table 4 instead lists the inner slopes $\gamma_{\rm in}$ from equation (22), with bootstrap errors, and the local slope $\gamma(R_{200m})$ of the integrated models.

\begin{table}
    \centering
    \begin{tabular}{c c c c}
    \hline
    Form & $\gamma_{\rm in}$ & $\gamma(R_{200m})$ \\
    \hline
    NFW & $1$ & $2.88$  \\
    1 & $0.917\pm0.012$ & $3.22$  \\
    2 & $0.841\pm0.033$ & $3.14$  \\
    3 & $0.874\pm0.022$ & $3.34$  \\
    % haloes (median) & -- & $\ldots$ & -- \\
    \hline
    \end{tabular}
    \caption{Logarithmic density slopes implied by the forms of Table~\ref{tab:generalised}: $\gamma_{\rm in}$ (extrapolated cusp) and $\gamma(R_{200m})$ (local slope of the integrated model).}
    \label{tab:slopes}
  \end{table}

\begin{figure*}
    \centering
    \includegraphics[width=\textwidth]{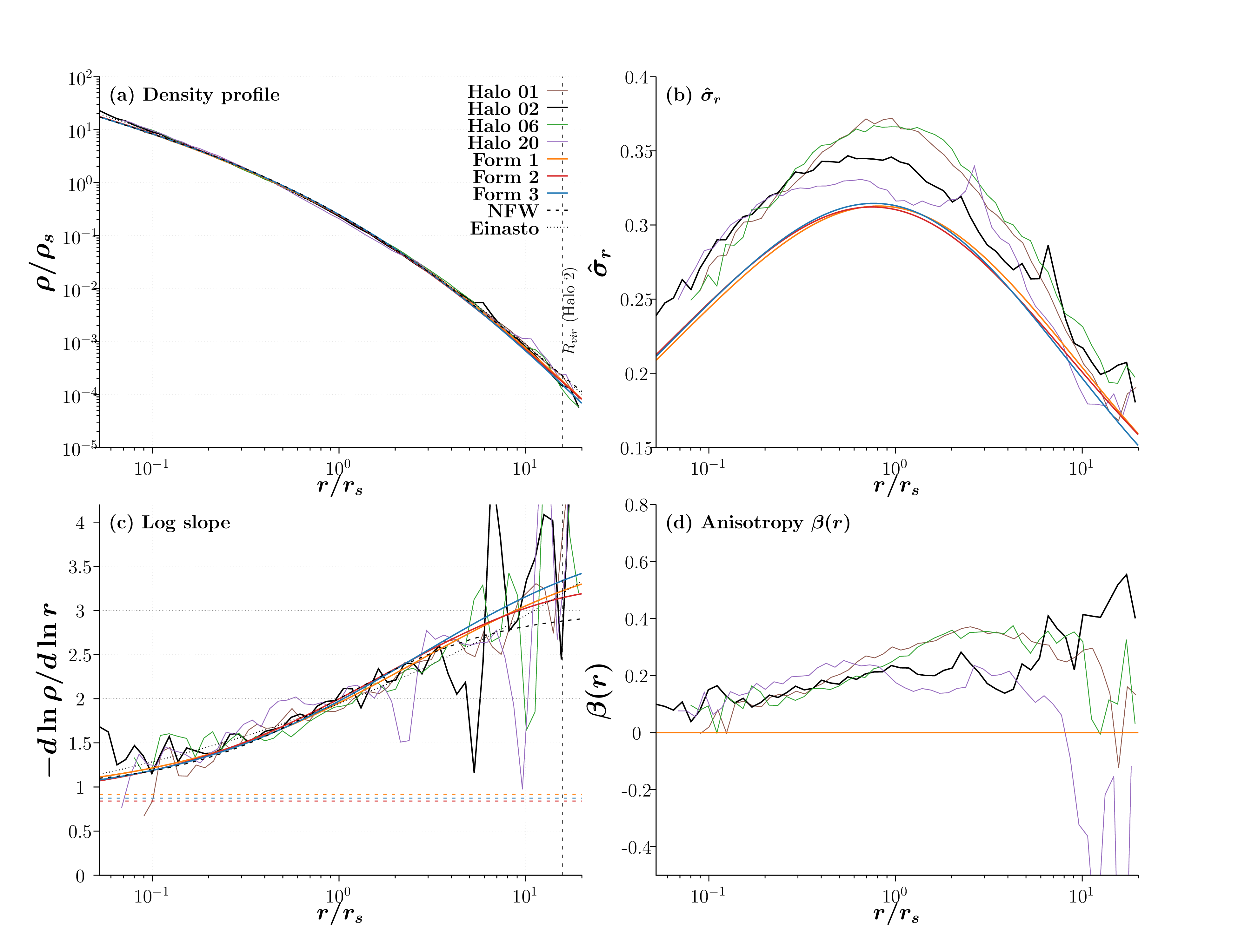}
    \caption{An illustration of the  implied observables under our three symbolic forms as well as their empirical values for Halo 01, Halo 02, Halo 06 and Halo 20. (a) A plot of the self-consistent density implied by forms 1,2 and 3 above. (b) The self-consistent dimensionless radial velocity dispersion implied by forms 1,2 and 3 above. (c) The slope of the logarithm of the density function. (d) Velocity anisotropy $\beta(r)$ of the haloes; the forms have $\beta = 0$ by construction. \textit{Orange Solid Line}: Self-consistent singular density implied by form 1. \textit{Red Solid Line}: Self-consistent singular density implied by form 2. 
    \textit{Blue Solid Line}: Self-consistent singular density implied by form 3. \textit{Black Dotted Line}: Einasto profile fit to the density of Halo 02 ($\alpha = 0.18$; this differs from the value in Table~\ref{tab:generalised}, which is fitted to the median DF). \textit{Black Dashed Line}: NFW profile of Halo 02. \textit{Dashed Orange/Red/Blue Line:} The inward asymptotic slopes of the respective models.  \textit{Black Solid Line:} Halo 02's associated empirical quantities for its bound set.}
    \label{fig:physical}
\end{figure*}
Fig. \ref{fig:physical}a implies that, within $R_{\text{vir}}$, our symbolic models generate a density that agrees excellently with both the NFW and Einasto profiles. In fact, they appear to compromise between both the Einasto and NFW profiles. Notice that Fig. \ref{fig:physical}c appears to show that the local slope steepens through $-2$ near $r_s$ and reaches $\gamma 
\approx 3.1$--$3.3$ at $R_{\rm vir}$. That is slightly steeper than NFW at the same radius. Beyond $R_{\rm vir}$ the
profiles are unconstrained extrapolations. The cusp solution generically truncates at a finite radius, and reaches the finite-mass Keplerian branch only at a critical exponent (Appendix~B).  The inner slopes agree with that of NFW to within $8$--$16\%$, but are all shallower. We argue this may partly reflect a consequence of gravitational softening. Our inner grid extends below the softening length, so the high-$\hat \varepsilon$ tail of the empirical DF samples a potential well that is shallower than a true cusp.

We also evaluate the associated radial velocity dispersion for all three analytic forms. The implied velocity dispersion can be derived from the spherical Jeans equation so that using our dimensionless variables
\begin{equation}
    \hat\sigma_r^{2}(\hat\Psi)\,\hat\rho(\hat\Psi) = \frac{8\sqrt{2}\pi}{3}\,
    q^{3/2}
    \int_{0}^{\hat\Psi}d\hat\varepsilon f(\hat\varepsilon)\,(\hat\Psi-\hat\varepsilon)^{3/2}\,
     ,
  \end{equation}
  where $\hat\sigma_r^2 = \sigma_r^2/\Psi(0)$ is the dimensionless radial
  velocity dispersion. For the functional form
  $f(\hat\varepsilon) \propto
  \hat\varepsilon^{\nu}/(1-\hat\varepsilon)^{\mu}$,
  a change of variables and \eqref{denspotdf} give
  \begin{equation}
    \hat\sigma_r^2 = \frac{\hat\Psi}{\nu+5/2}\,
    \frac{{}_2F_1(\nu+1,\mu;\nu+7/2;\hat\Psi)}{{}_2F_1(\nu+1,\mu;\nu+5/2;\hat\Psi)},
  \end{equation}
  where, by isotropy, $\sigma_r^2 = \sigma_\theta^2 = \sigma_\phi^2$. Fig. \ref{fig:physical}b and \ref{fig:physical}d show the main limitation of an energy-only DF. It has $\beta = 0$ by construction, whereas the haloes are radially biased, with median $\beta \simeq 0.15$--$0.25$ within $R_{200m}$. The implied dispersion profiles reproduce the shape of $\hat\sigma_r$, rising to a maximum near $r_s$ and declining outwards, but lie below the measured profiles: by 7--14\% (median over $0.05r_s$--$20r_s$), and by $\sim15\%$ at $r_s$ for haloes 01 and 06. This offset is accounted for by the radial anisotropy of the haloes. The measured $\rho$, $\sigma_r$, $\beta$ and $\Psi$ satisfy the anisotropic Jeans equation to within 1\%, so the bound-particle potential is consistent with the particle velocities. Solving the same equation with $\beta = 0$, i.e. for an isotropic halo with identical $\rho$ and $\Psi$, gives $\sigma_r$ 9--13\% below the measured profiles; integrating each halo's own empirical $f(\varepsilon)$ gives 8--13\%, and the symbolic forms agree with each halo's own $f(\varepsilon)$ to within 1\% and with the isotropic Jeans baseline to within 2\%. The offset is therefore a property of any energy-only DF rather than of the fitted forms. The isotropic dispersion at the same radius, $\sqrt{\langle v^2\rangle/3} = \sigma_r\sqrt{1 - 2\beta/3}$, accounts for only 5--9\%, because the $2\beta/r$ term in the Jeans equation accumulates over all radii outside $r$. Capturing the velocity structure requires an anisotropic DF, $f(E,L)$,
consistent with \citet{Kumar2026}, who find $f(E,L)$ statistically preferred
over $f(E)$ for Auriga haloes. See the Appendix for the numerical procedure used to solve Poisson's equation and for the asymptotic behaviour of the models.

  All in all, our expressions are the generic asymptotic skeleton of any isotropic distribution function whose density has power-law limits; PySR has effectively \enquote{recovered} the limits and \enquote{patched} them together. The non-trivial result is therefore that the Auriga cosmological haloes require little correction to it when preprocessed under this pipeline.

\subsection{Sources of Residual Error}
\label{sec:residual}
\begin{figure*}
    \centering
    \includegraphics[width=\textwidth]{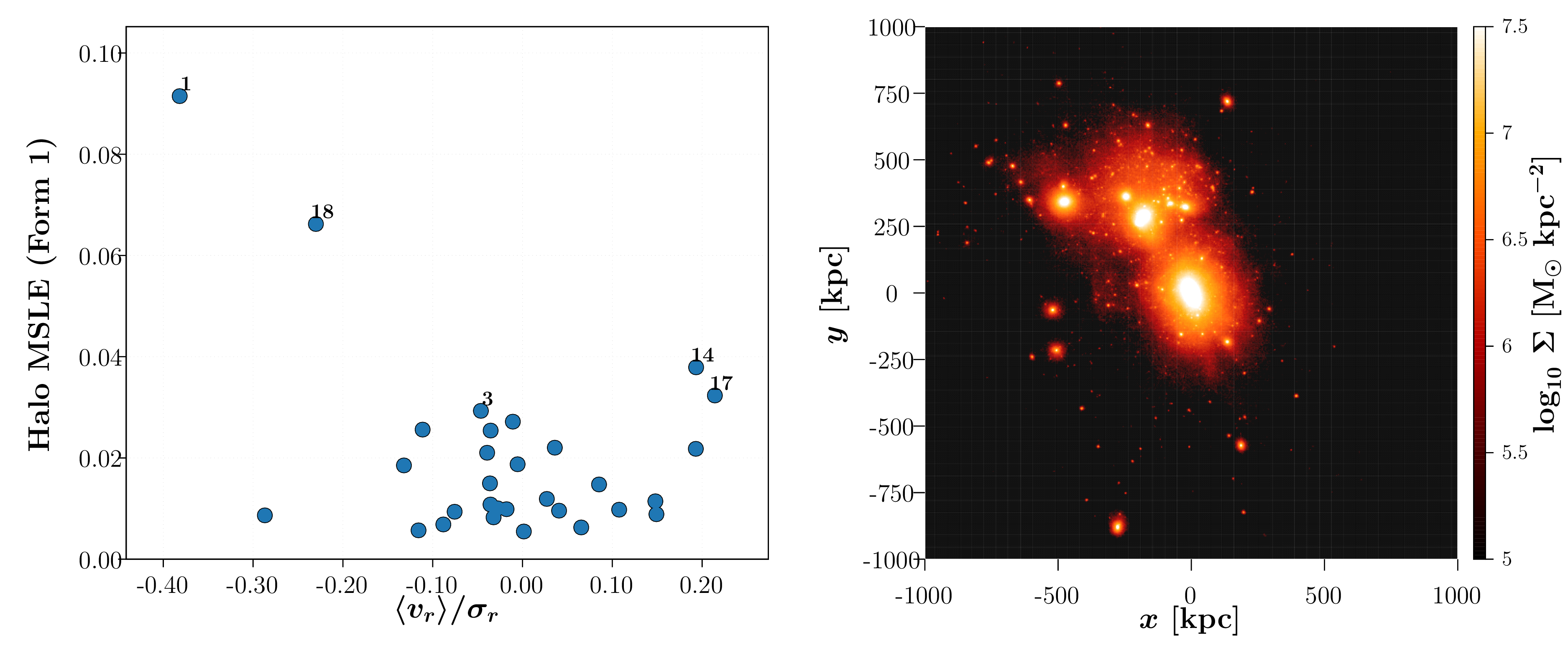}
    \caption{An illustration probing the effect of proper radial motion as a source of residual error. (a) A scatter plot between the MSLE achieved by form 1 for a given halo's distribution function, against a measure of radial motion. (b) The projected 2D density maps of Halo 01's bound subset of particles.}
    \label{fig:error}
\end{figure*}

Whilst the issue of finding analytic expressions for our dataset is inevitably degenerate, uncertainty and noise are the distinguishing factors that render differentiating between two simple parametric expressions difficult. As stated above, this \textit{resolution} is the accuracy of the median of the similarity transformed curves' loss when scaled back in raw coordinates. From a physical standpoint, we can expect error to arise from three sources present in our bound halo structures: non-equilibria, anisotropy and asphericity. Upon inspection of individual haloes, the largest single contributor to the average MSLE loss of the ensemble is Halo 01. Upon closer inspection of the bound subset selected, one notices that Halo 01 is clearly in disequilibrium and undergoing a merger between two large central subhaloes, as illustrated in Fig. \ref{fig:error}b. An excellent proxy to quantify this disequilibrium consists of \(\langle v_{r} \rangle/\sigma_{r}\), a measure of normalised proper residual radial motion. We use proper velocities here since these are the velocities used for classification, dynamic selection with Hubble flow. In perfect equilibrium, the mean radial velocity will vanish and hence \(\langle v_{r} \rangle/\sigma_{r}\) will be a strong proxy for disequilibrium.  
Fig. \ref{fig:error}a provides a scatter plot of this proxy against the MSLE loss for our leading expression across the haloes.   In particular, the worst fitting halo, Halo 01, is associated with a merger and exhibits elevated radial motion, while the remaining haloes collapse onto a broadly common relation despite variations in shape and anisotropy. However, it is unclear whether radial motion is correlated with the loss in the remaining ensemble of haloes. We note similarly that our models perform particularly poorly on Halo 18, as illustrated in Fig. \ref{fig:error}a. We note these are the haloes with a particularly large $q$. 

Overall, it is difficult to conclude what the leading sources of uncertainty were. Whilst Halo 01 may easily be dismissed as anomalous, the remaining haloes appear to illustrate no significant trends with respect to disequilibria, asphericity and anisotropy. The Spearman rank correlations between the MSLE loss and each of these
proxies, $\langle v_r\rangle/\sigma_r$ (signed and in magnitude), the anisotropy parameter $\beta$ and the minor-to-major axis ratio $c/a$, are not significant at the 5 per cent level ($N=30$). However, we have used proxies as measures of these three variables and hence more accurate and extensive work must be done on understanding the exact causes of these losses. One could explain these findings through a complex dependence on all three factors, that cannot directly be probed on a single scatter plot and could be addressed by computing partial correlations. Otherwise, since the outer tail of these distribution functions is sensitive to the outskirts of haloes, these proxy results may simply indicate that the dynamics in the outer halo are halo dependent, not directly associated to global dynamic measures, such as merger or accretion history, especially since we select bound particles using historic trajectories. The dynamic selection and the restriction to $\hat\varepsilon \geq 0.03$ reduce subhalo structures and much of the fly-by material. They do not remove the radial anisotropy of the haloes, $\beta \approx 0.2$, so the energy-only DF describes their density profiles well but not their velocity structure as seen in Fig. \ref{fig:physical}b.   

\subsection{Conclusion}

In this work we have developed a pipeline to extract approximate empirical distribution functions from cosmological simulations as well as providing the simplest, best fitting physical expressions associated with this data. In fact, we found an extremely simple functional form \(C\hat \varepsilon^{\nu}/(1 - \hat \varepsilon)^{\mu}\) for DMO halo distribution functions whose simplest member closely agrees with NFW and Einasto fits within the virial radius, out of a search over $\sim 500$ candidates. Its simplest member, Form~1, is available in closed form, satisfies the
sufficient condition for stability by construction, and is as accurate as the Einasto DF with the same number of free parameters. Its central cusp is slightly shallower than NFW, plausibly a consequence of gravitational softening. The implied density steepens to $\gamma \simeq 3.1$--$3.4$ at $R_{\rm vir}$, slightly steeper than NFW. This rests on the weakly bound part of the DF, which the data constrain least. 

This work has several natural extensions. Notably, we could extend the logic employed to regress two parameter functions $f(E,L)$  that can account for anisotropy by taking angular momentum as an argument. Also, whilst we have stayed clear of the effect of Hubble flow in our distribution functions, its role in the energy-space DFs of halo outskirts appears largely unexplored, lying between the scales of cosmology and astrophysics. One could imagine  Hubble flow introducing both time dependence and reference frame complications.

\section*{Acknowledgements}
We acknowledge use of the Auriga cosmological simulations \cite{AurigaWebsite} and the PySR symbolic regression package \cite{Cranmer2023PySR}. We thank members of the Institute of Astronomy dynamics group for helpful discussions.
\section*{Data Availability}

The code is available on GitHub at \url{https://github.com/mbotbol2/Symbolic-Regression-of-Auriga-DFs}.

\bibliographystyle{mnras}
\bibliography{example} 

\appendix

\section{PySR \& Hyperparameters Used}
PySR is a machine learning algorithm, developed by Miles Cranmer in \cite{Cranmer2023PySR}, that performs symbolic regression on a given dataset. As discussed above, PySR is implemented in Julia, with a Python frontend, and represents analytic expressions with trees. Using tournament selection, PySR evolves populations of analytic expressions and updates the Pareto front throughout the search. There are several genetic algorithms it employs, including \enquote{mutation}, \enquote{crossover},
\enquote{simplification}, and \enquote{constant optimisation}. The algorithm is far more efficient than exhaustive symbolic regression and genetically evolves several populations of operators independently so as to maximise the explored search space. We refer the reader to the paper for more details.

As mentioned in the main body of this work, the runs were performed using the set of unary operators $\{ \log, \exp, \sqrt, \arcsin, \sinh \}$. For each run, the Pareto front appeared to converge after approximately $2000$ iterations. We set the nesting constraints so each operator could nest any other of the five operators at most once, and never itself, so as to prevent unphysical complex nested expressions. The maximum complexity of our operators was set to $40$, however since we regressed transformed variables, we inevitably generated rather complex expressions of greater complexity. We set the number of independently evolving populations to $36$ and followed the advice provided in \cite{pysr_tuning} for the remaining fine tuning of hyperparameters. 

\section{SOLUTION BRANCHES OF POISSON'S EQUATION AND ASYMPTOTIC SCALING}
In the main body of this work, we consider the general function form $f(\hat \varepsilon) = C\hat \varepsilon^{\nu}/(1-\hat \varepsilon)^{\mu}$ to derive results for our symbolic fits. Here, we derive the asymptotic properties of the radial density implied by this general distribution function. First, let us quickly derive the isotropic density–potential relation \eqref{denspotdf}. We have 
\[\hat\rho(\hat\Psi) = 4\sqrt{2}\pi\, q^{3/2}\int_0^{\hat\Psi}d\hat\varepsilon
  f(\hat\varepsilon)\sqrt{\hat\Psi-\hat\varepsilon}\, 
  \]
  \[= 4\sqrt{2}\pi C q^{3/2}\int_0^{\hat\Psi} d\hat\varepsilon\frac{\hat\varepsilon^{\nu}}{(1-\hat\varepsilon)^{\mu}}\sqrt{\hat\Psi-\hat\varepsilon}\,\] 
so that by setting $t =  \hat \varepsilon/\hat \Psi$, we have 
\[\hat\rho(\hat\Psi) = 4\sqrt{2}\pi C q^{3/2}\,\hat\Psi^{\nu+3/2}\int_0^1dt\cdot 
  t^{\nu}(1-t)^{1/2}(1-\hat\Psi t)^{-\mu}\,\]
One can notice this is the Euler representation of the hypergeometric function and thus express this more compactly as \eqref{denspotdf}.

It is useful to define $x = \sqrt{k}\hat r$ with $k = 16\sqrt2\,\pi^2Cq^{1/2}\mathcal{B}(\nu+1,\frac{3}{2})$ as well as $F={}_2F_1\!\left(\nu+1,\mu;\nu+\frac52;\hat\Psi\right)$, so that Poisson's equation reduces to 
\begin{equation}
  \hat \Psi''+  \frac{2}{x}\hat\Psi'=-\hat\Psi^{n}F(\hat\Psi),
  \qquad \hat \Psi(0)=1, \qquad n= \nu +3/2
  \label{eq:B_poisson}
\end{equation}
We first address the inner slope. Notice, as $\hat \varepsilon \to 1$, we have
$f\propto(1-\hat\varepsilon)^{-\mu}$ which, with a change of variables inside
\eqref{denspotdf}, implies $\hat\rho\propto(1-\hat\Psi)^{-(\mu-\frac{3}{2})}$,
divergent for $\mu>3/2$. Since a cusp $\hat\rho\propto x^{-\gamma_{\rm in}}$ encloses
$M\propto x^{3-\gamma_{\rm in}}$, $\hat\Psi'=-M/x^2$ implies
$1-\hat\Psi\propto x^{2-\gamma_{\rm in}}$. Hence, matching exponents gives
\begin{equation}
  \gamma_{\rm in}=2\left(\,\frac{\mu-3/2}{\mu-1/2}\right).
\end{equation}
Near the cusp we therefore have $1-\hat\Psi = c\,x^{s}$ with $s=2-\gamma_{\rm in}$,
where the constant $c$ is fixed by Poisson's equation. Together with
$\hat\Psi(0)=1$, this determines the solution uniquely for given $(\nu,\mu)$,
and supplies the initial conditions $\hat\Psi(x_0)$ and $\hat\Psi'(x_0)$ used in
the numerical scheme.

We now turn to the outer region. There are two cases to distinguish, $\hat\Psi$ either vanishes at a finite radius or tends to zero. In the latter case, as $\hat \Psi \to 0$, $F(\hat\Psi) \to 1$. Thus, Poisson's equation reduces to the Lane-Emden equation of index $n$, namely  \begin{equation}
  \hat \Psi''+  \frac{2}{x}\hat\Psi'=-\hat\Psi^{n}
  \label{eq:B_poisson_limit}
\end{equation}
To remove $x$ dependence, we use the change of variable $\hat \Psi = x^{-\lambda}w(\ln(x))$, $\lambda = 2/(n-1)$, we have 
\begin{equation}
\ddot{w} + (1 - 2\lambda) \dot{w} - \lambda(1 - \lambda) w + w^{n} = 0
\end{equation}
Its two equilibria are the power-law solutions. The origin, $w=0$, corresponds
to the \textit{Keplerian solution}, with $\hat \Psi \propto x^{-1}$ and with
\begin{equation}
\gamma_{\rm out} = \nu+3/2
\end{equation}
The non-trivial equilibrium,
$w_{\star}=[\lambda(1-\lambda)]^{1/(n-1)}$, corresponds to the \textit{scale-free solution}, implying
$\hat\Psi\propto x^{-\lambda}$, 

\begin{equation}
\gamma_{\rm out}=2n/(n-1) = 2\left(\frac{\nu +3/2}{\nu +1/2}\right)
\end{equation}

as well as an infinite mass system. For $3<n<5$ the friction term $(n-5)/(n-1)$ is negative, so solutions gain
energy and spiral away from $w_{\star}$ with growing amplitude, until $w$, and hence
$\hat\Psi$, vanishes at finite radius. The exception is the cusp solution that lands on the saddle $w=0$ (the
Keplerian branch), which for $\mu=\nu$ occurs at $\nu_c\simeq2.336$, independent of $C$ and $q$. Note that the oscillations in log slope of our numerical solution represents the orbit of the solution in phase space. 

\section{Numerics for the Singular Solution of Poisson's Equation}

In the main body of this work, we considered the general function form $f(\hat \varepsilon) = C\hat \varepsilon^{\nu}/(1-\hat \varepsilon)^{\mu}$ to derive results for our symbolic fits. Here, we discuss the numerics employed to solve for the singular solution of Poisson's equation. Using the dimensionless variables $(x, \hat \rho, \hat \Psi)$, we have seen that integrating over velocities yields the following isotropic density–potential relation
\[\hat\rho(\hat \Psi) = 4\sqrt{2}\pi C q^{3/2}\hat
  \Psi^{\nu+3/2}\mathcal{B}(\nu+1,
  3/2){}_2F_{1}(\nu+1,\mu;\nu+5/2;\hat \Psi)\]
Next, it suffices to solve the dimensionless Poisson equation \eqref{eq:B_poisson},
\[\frac{d^2\hat\Psi}{dx^2} + \frac{2}{x}\frac{d\hat\Psi}{dx} =
  -\hat\Psi^{n}F(\hat\Psi)\]
where we use Halo 02 scale length and density to form dimensionless variables. Now, this second order ODE may be expressed as a system of two first order ODEs, by introducing $t = \ln(x)$ and $u = d\hat\Psi/dt$. This yields
\begin{equation}
\frac{d\hat\Psi}{dt} = u, \qquad \frac{du}{dt} = -u -
  \hat\Psi^{n}F(\hat\Psi)\,e^{2t}
\end{equation}
Now, the singular solution corresponds to that with $(\hat \Psi(0) = 1, \hat \rho(0) = \infty)$. We have seen that as $x \to 0$, $\hat \Psi(x) \sim 1 - c\, x^{s}$, with $s = 2/(\mu-1/2)$ and $c$ known. In practice, we set initial conditions in $\hat \Psi, \hat \Psi^{'}$ at $x = x_0$, for $x_0$ small, and integrate outwards.

\bsp	
\label{lastpage}
\end{document}